\documentclass[%
 amsmath,amssymb,
 aps,
 prfluids,
]{revtex4-2}

\usepackage[notes]{ctmmath-v3_revtex} 

\usepackage{subcaption}
\usepackage{graphicx}% Include figure files
\usepackage{dcolumn}% Align table columns on decimal point

\usepackage{bm}% bold math
\begin{document}

\preprint{APS/123-QED}

\title{Non-Newtonian fluid flow in porous media}

\author{Christopher A. Bowers}
\email{bowersca@live.unc.edu}
\author{Cass T. Miller}
\affiliation{%
 Department of Environmental Sciences and Engineering, University of North Carolina at Chapel Hill, North Carolina 27599, USA
}%

\date{\today}

\begin{abstract}
Single-fluid-phase porous medium systems are typically modeled at an averaged length scale termed the macroscale, and Darcy's law is typically relied upon as an approximation of the momentum equation under laminar flow conditions. Standard approaches for modeling macroscale single-fluid-phase flow of non-Newtonian fluids extend the standard Newtonian model based upon Darcy's law using an effective viscosity and assuming that the intrinsic permeability is invariant with respect to fluid properties. This approach results in a need to determine the effective viscosity for every fluid composition and flow rate. We use the thermodynamically constrained averaging theory (TCAT) to examine the formulation and closure of a macroscale model for non-Netwonian flow that is consistent with microscale conservation principles and the second law of thermodynamics. A direct connection between microscale and macroscale quantities is used to formulate an expression for interphase momentum transfer for non-Newtonian flow in porous medium systems. Darcy's law is shown to approximate momentum transfer from the fluid phase to the solid phase. This momentum transfer is found to depend on the viscosity at the solid surface, which is only invariant for Newtonian flow. As a consequence of the derived equation for momentum transfer, the commonly called intrinsic permeability is not invariant for non-Newtonian flow, which is an assumption that underlies standard effective viscosity approaches for modeling non-Newtonian flow. TCAT is used to derive a macroscale equation relating the flow rate and the pressure gradient and dependent upon fluid properties. Non-Newtonian flow can be modeled using this equation along with characterization of the fluid properties and medium characterization under a single flow condition, breaking the need to investigate all flow and composition conditions that are hallmarks of extant approaches. This new approach is validated for model systems and used to interpret results from the literature, including an evaluation of conditions under which a transition occurs away from strictly laminar flow conditions. The results from this work form a basis for more rigorous and realistic modeling of non-Newtonian flow in porous medium systems.
\end{abstract}

\maketitle

\section{Introduction}

Most fluids encountered in natural and industrial applications, with the exception of a few such as water or air, are non-Newtonian in character. Non-Newtonian fluids flow in complex ways that are not consistent with Newton's law of viscosity, i.e. their viscosity changes as function of shear rate, all other things being equal. The complexity of non-Newtonian versus Newtonian fluid mechanics in porous medium systems has lead to a misunderstanding of physical data across many fields. It is typically assumed that macroscale models that were derived for Newtonian fluids can be used to model non-Newtonian flows using simple {\em ad hoc} extensions such as effective viscosities. This misunderstanding carries widespread implications, as non-Newtonian fluids appear regularly in biofluidics \cite{Peyrounette_Davit_etal_18,Bessonov_Simakov_etal_16,Sriram_Intaglietta_etal_14,Chakraborty_05,Rabby_Razzak_etal_13}, geophysics \cite{Gerrard_Perutz_etal_52,Ancey_Meunier_04,Ancey_07,Barr_Pappalardo_etal_04,Hulme_74,Wu_02,Sonder_Zimanowski_etal_06,Lev_Spiegelman_etal_12}, and subsurface processes such as hydraulic fracturing and enhanced oil recovery \cite{Osiptsov_17,Norman_Jasinski_etal_96,Barati_Liang_14,Zhang_Prodanovic_etal_19,Barbati_Desroches_etal_16,Xie_Lv_etal_18,Perrin_Tardy_etal_06,Buchley_Lord_73}. 

While it has often been assumed that models developed for Newtonian fluids would be adequate for predicting non-Newtonian fluid flow, it has been found in many cases that the non-Newtonian rheology has a major impact on flow characteristics \cite{Chen_Lu_06,Sochi_10,Mejia_Mongrain_etal_11,Billen_Hirth_05,Skelland_67,Mendes_07,Thompson_Soares_16}. The disparity between Newtonian and non-Newtonian fluid mechanics is due to the viscosity changing throughout the domain and with flow conditions for a non-Newtonian fluid, as opposed to the entirety of the fluid being the same viscosity for Newtonian fluids. This complication in the determination of flow becomes apparent when modeling flow through a capillary tube. While laminar, non-Newtonian flow in a capillary tube may be calculated when only knowing the shear stress at the wall, assuming that the shear stress at the centrum is zero, the viscosity model chosen affects the flow rate in the system, which is calculated by integrating over the range of shear stresses \cite{Bird_Stewart_etal_02,Skelland_67,Sochi_15,Perrin_Tardy_etal_06}. When moving to more complex porous medium systems, the significance of such non-Newtonian behavior increases, and the assumption that Newtonian models will suffice may lead to an incorrect prediction of flow rates for a given applied force.

The current state of the science for non-Newtonian flow modeling in porous media is highly empirical, requiring experimentation or modeling results for every non-Newtonian fluid of interest, flowing within every geometry of interest over a wide range of flow rates to produce statistically fit parameters to be predictive \cite{Hayes_Afacan_etal_96,Liu_Masliyah_99,Zami_Loubens_etal_16,Zhang_Prodanovic_etal_19}. Models used are based on Darcy's law, assuming that the hydraulic conductivity of the system can be broken apart into a geometric intrinsic permeability term and a viscosity term, which in the case of non-Newtonian fluid flow is called the effective viscosity \cite{Sadowshi_Bird_65,Sorbie_Clifford_etal_89,Chauveteau_82,Parnas_Cohen_87,Hayes_Afacan_etal_96,Liu_Masliyah_99,Perrin_Tardy_etal_06,Tosco_Marchisio_etal_13,Castro_Radilla_17,Berg_Wunnik_17,Zamani_Bondino_etal_17,Castro_19,Zhang_Prodanovic_etal_19}. Due to the {\em ad hoc} nature of the effective viscosity assumption, there is confusion as to what such a parameter means physically, with some suggesting that it represents the viscosity at the fluid-solid interface \cite{Sadowshi_Bird_65,Sorbie_Clifford_etal_89}, that it is the average viscosity throughout the fluid \cite{Bird_Stewart_etal_02,Perrin_Tardy_etal_06,Zamani_Bondino_etal_17}, that it is a ``fictitious" viscosity \cite{Hayes_Afacan_etal_96,Tosco_Marchisio_etal_13,Zhang_Prodanovic_etal_19}, and that it is a ``macroscopic" viscosity \cite{Liu_Masliyah_99}. Without consensus on the fluid parameters that are significant, there has been little progress in theoretical understanding of what is occurring during non-Newtonian flow in porous media. What is required is a new theory of macroscopic, non-Newtonian flow that is rigorously formulated from first principles.  

The terminology of length scales is inconsistent in the porous medium physics literature. We will rely on two scales, the microscale, or pore scale, and the macroscale, or porous medium continuum scale. The microscale is a scale at which the laws of continuum mechanics are applicable but the distribution of all phases is resolved in space and time. The macroscale is a scale where a point represents a centroid of an averaging region containing all phases in the system.  The macroscale is the scale at which most applications must be described because of computational limitations.  There have been several successful attempts to derive the equations of flow by averaging to the macroscale from the microscale \cite{Hubbert_56,Gray_ONeill_76,Neuman_77,Whitaker_86,Nordbotten_Celia_etal_07,Gray_Miller_etal_13}, and even some attempts when the fluid is non-Newtonian \cite{Hayes_Afacan_etal_96,Liu_Masliyah_99}; however, it is usually the case that the hydraulic conductivity of the system is what is derived by averaging, and it is assumed that this hydraulic conductivity may be decomposed into an intrinsic permeability and a fluid viscosity after averaging. Even in the case where it is acknowledged that a non-Newtonian viscosity may enter averaging \cite{Gray_ONeill_76}, or that the actual permeability is different than the intrinsic permeability \cite{,Hayes_Afacan_etal_96,Liu_Masliyah_99}, the actual microscale variables that must be averaged to generate a system permeability and a viscosity have not been investigated. The thermodynamically constrained averaging theory (TCAT) is a model building approach that can be used to derive macroscale models based upon microscale principles \cite{Gray_Miller_06,Gray_Miller_etal_13,Gray_Miller_14}. TCAT has been used to derive model hierarchies for a wide variety of systems including single-fluid porous medium systems \cite{Gray_Miller_06,Miller_Gray_08,Gray_Miller_09,Gray_Miller_09c,Gray_Miller_14}, two-fluid porous medium systems \cite{Jackson_Miller_etal_09,Gray_Miller_14,Gray_Dye_etal_15,Rybak_Gray_etal_15}, sediment transport in surface waters \cite{Miller_Gray_etal_19}, and tumor growth \cite{Sciume_Shelton_etal_12,Sciume_Shelton_etal_13}. However, TCAT methods have not been used to formulate macroscale models for non-Newtonian fluid flow in porous medium systems. 

The goal of this paper is to develop and evaluate a new theory for the laminar flow of non-Newtonian fluids in a single-fluid porous medium that is consistent with microscale conservation and thermodynamic principles. The specific objectives of this work are (1) to examine an existing model hierarchy for single-fluid-phase flow that can be used to model non-Newtonian flow; (2) to derive a momentum transfer approximation that is applicable to non-Newtonian fluids; (3) to evaluate and validate the resultant model by comparison to ideal systems; and (4) to use the derived model to analyze available data, including flows that are transitioning away from a pure laminar state.

\section{Computational Methods}

To assist in a fundamental investigation of macroscale modeling of non-Newtonian flow through a porous medium, highly resolved microscale simulations were performed. In particular, three different media were investigated: a set of slits through which a non-Newtonian fluid was flowing in a parallel arrangement; randomly packed spheres with flow in one direction; and ellipsoids oriented with flow aligned with the principal directions of the media.  These simulations had two fundamental components: medium generation, and flow simulation, which are described in turn. 

In the case of parallel slits, the medium was comprised of three slits of differing aperature, but which had the same length and width. Calculation of the flow rate for each slit was carried out using an existing semi-analytical solution \cite{Sochi_15} for Cross model fluids. Matlab was used to numerically solve for the shear rates occurring at the wall of the slits and to solve for the hypergeometric function, ${}_2F_1$, which is included in the semi-analytical solution. The \texttt{vpasolve} function within Matlab was used to solve the Cross model equation 
\begin{equation}
 \lrb{ \hat{\mu}_\infty + \frac{\hat{\mu}_0 - \hat{\mu}_\infty}{1 + (m\dot{\gamma_w})^n} }\dot{\gamma_w} = \tau_w\;,
\end{equation}
for the wall shear rate $\dot{\gkg}_w$ for a given wall shear stress $\tau_w$.

For the packed sphere medium, the location of each sphere center and its radius were generated using an algorithm based on \cite{Williams_Philipse_03}, packing the spheres randomly with log-normally distributed radii. These randomly packed sphere locations and radii were input in OpenFOAM, using the searchableSphere utility, and this was used to generate a castellated medium domain using the snappyHexMesh utility \cite{Greenshields_18}. Non-Newtonian flow simulations were carried out using the non-Newtonian solver built into the simpleFoam utility \cite{Greenshields_18}, and simulation post-processing was carried out using Paraview to calculate the variables of interest \cite{Ayachit_19}. The simpleFoam utility solves the mass and momentum balance equations
\begin{gather}
\del\cdot\vec u, \\
\del\cdot\lrp{\vec u \vec u } - \del\cdot\ten\tau = -\del p+ \vec S_u\;,
\end{gather}
where $\vec u$ is the microscale velocity vector within a cell in the system, $\ten\tau$ is the stres tensor, $p$ is the microscale pressure, and $\vec S_u$ is a momentum source term \cite{Greenshields_18}. The SIMPLE algorithm is used to solve these equations \cite{Caretto_Gosman_etal_72,Greenshields_18}. The incompressible stress tensor used in OpenFOAM is
\begin{equation}
\ten\tau = 2\Hat{\mu}\ten d\;,
\end{equation}
where $\ten d$ is the rate of strain tensor, defined as
\begin{equation}
\ten d = \frac 12 \lrb{\del\vec u + \lrp{\del\vec u}\T }\;.
\end{equation}
The shear rate convention used in OpenFOAM is 
\begin{gather}
\dot{\ten{\gkg}} = 2\ten d \\
\dot{\gkg} = |\dot{\ten{\gkg}}| = \sqrt{\dot{\ten{\gkg}}\dd\dot{\ten{\gkg}}}\;,
\end{gather}
where $\dot{\gkg}$ is the shear rate used in the Cross model viscosity equation. In OpenFOAM, the Cross model function is defined as
\begin{equation}
\hat{\mu}_\infty + \frac{\hat{\mu}_0 - \hat{\mu}_\infty}{1 + (m\dot{\gamma})^n}\;.
\end{equation}
OpenFOAM uses a finite volume method to solve the microscale conservation equations for individual volume cells throughout the domain. Flow rates were calculated using the flowRatePatch utility to integrate over the inlet and outlet of the system. Pressure boundary conditions were applied to the inlet and the outlet of the system, while periodic boundary conditions were imposed on all other sides of the square domain. 

For the system of ellipsoids an STL mesh file was generated using FreeCAD, an open-source 3D parametric modeler, which was then meshed using the snappyHexMesh utility in OpenFOAM. In this case none of the ellipsoids were touching, and the "snapping" utility was used to generate a more accurate mesh \cite{Greenshields_18}. The simpleFoam utility was again used with pressure boundary conditions set on two opposed faces of the system, with cyclicAMI boundaries set on each other face. The cyclicAMI utility in OpenFOAM is used when cyclic mesh interfaces are not one-to-one, and an interpolation is carried out to match the two faces to one another \cite{Greenshields_18}. The faces for which pressure boundary conditions were chosen changed with each simulation to change the principal direction of flow. Post-processing averaging was carried out in Paraview as was done for the packed spheres case.

\section{Traditional Non-Newtonian Modeling}

Non-Newtonian fluids that do not follow Newton's law of viscosity, which is 
\begin{equation}
\tau = \hat{\mu}\dot{\gamma}\;,
\end{equation}
where $\tau$ is the shear stress, $\hat{\mu}$ is the dynamic viscosity of the fluid, and $\dot{\gamma}$ is the shear rate of the fluid. More explicitly, the dynamic viscosity of a non-Newtonian fluid changes as the shear stress applied to that fluid changes, all other things being equal. There are several different kinds of non-Newtonian fluids, such as those with time-dependence, viscoelastic fluids, and shear rate dependent fluids. Shear thinning fluids are especially important in industrial applications \cite{Osiptsov_17,Norman_Jasinski_etal_96,Barati_Liang_14,Zhang_Prodanovic_etal_19,Barbati_Desroches_etal_16,Xie_Lv_etal_18,Perrin_Tardy_etal_06,Buchley_Lord_73,Skelland_67,Mendes_07,Thompson_Soares_16,Sochi_15,Hayes_Afacan_etal_96,Parnas_Cohen_87,Boger_77,Subbaraman_Mashelkar_etal_71,Chhabra_Uhlherr_79} and geophysical flow of importance \cite{Gerrard_Perutz_etal_52,Ancey_Meunier_04,Ancey_07,Barr_Pappalardo_etal_04,Hulme_74,Wu_02,Sonder_Zimanowski_etal_06,Lev_Spiegelman_etal_12,Billen_Hirth_05}. At steady state, non-Newtonian fluids can generally be described by an equation of the form
\begin{equation}
\tau = \hat{\mu}(\dot{\gamma})\dot{\gamma}\;,
\end{equation}
where $\hat{\mu}(\dot{\gamma})$ indicates a dynamic viscosity that is a function of the shear rate. Various functional forms for the dynamic viscosity have been advanced \cite{Sochi_10,Bird_Stewart_etal_02}. 

The Cross model is a common non-Newtonian fluid model \cite{Cross_65,Sochi_10,Sochi_15,Skelland_67,Zhang_Prodanovic_etal_19}, which may be written as
\begin{equation} \label{eq:Cross_Model}
\hat{\mu}(\dot{\gamma}) = \hat{\mu}_\infty + \frac{\hat{\mu}_0 - \hat{\mu}_\infty}{1 + (m\dot{\gamma})^n}\;,
\end{equation}
where $\hat{\mu}_\infty$ is the infinite shear dynamic viscosity, $\hat{\mu}_0$ is the zero shear dynamic viscosity, $m$ is a measure of intramolecular attraction forces, and $n$ is a non-Newtonian behavior index \cite{Cross_65}. It is common to utilize a power law model, which does not have Newtonian limits, but it has been found that this model provides an inaccurate description of many fluids \cite{Boger_77,Subbaraman_Mashelkar_etal_71,Chhabra_Uhlherr_79}. 

\begin{figure*}[t!]
    \centering
    \begin{subfigure}[b]{0.49\linewidth}
        \includegraphics[width=\linewidth]{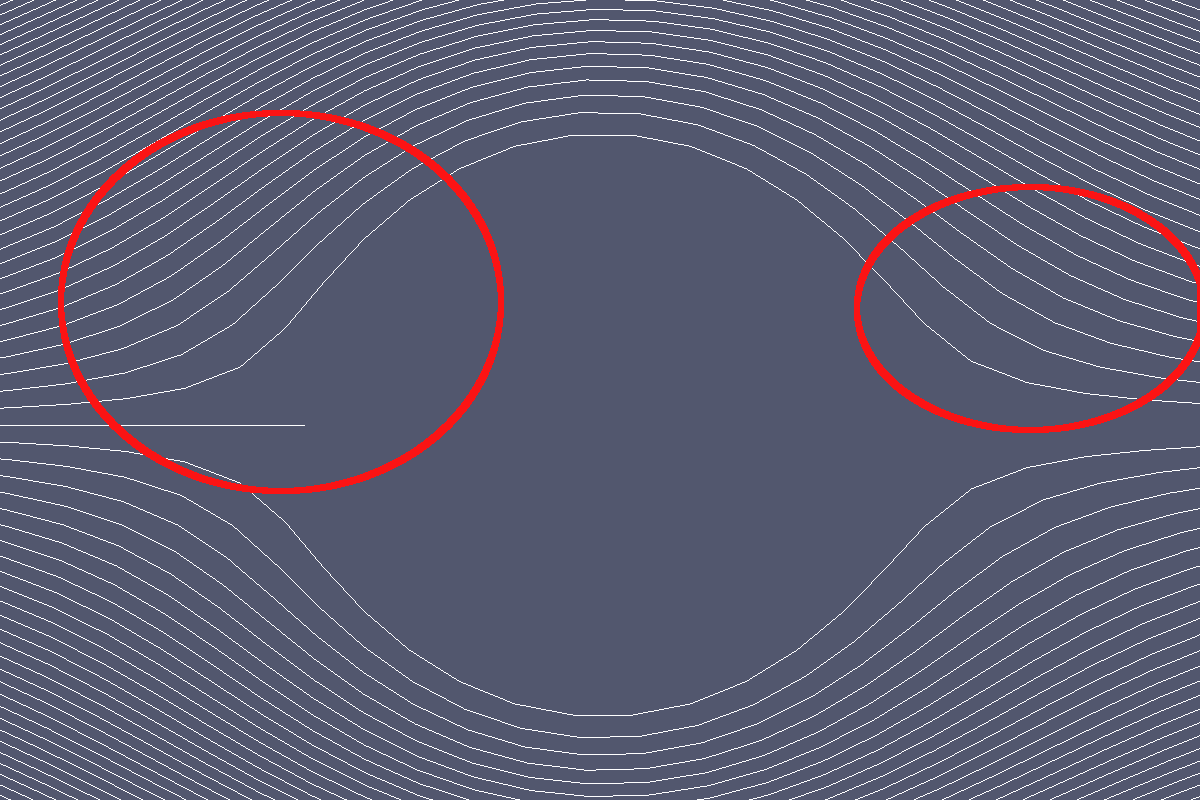}
        \caption{Streamlines for Newtonian fluid.}
    \end{subfigure}
    \begin{subfigure}[b]{0.49\linewidth}
        \includegraphics[width=\linewidth]{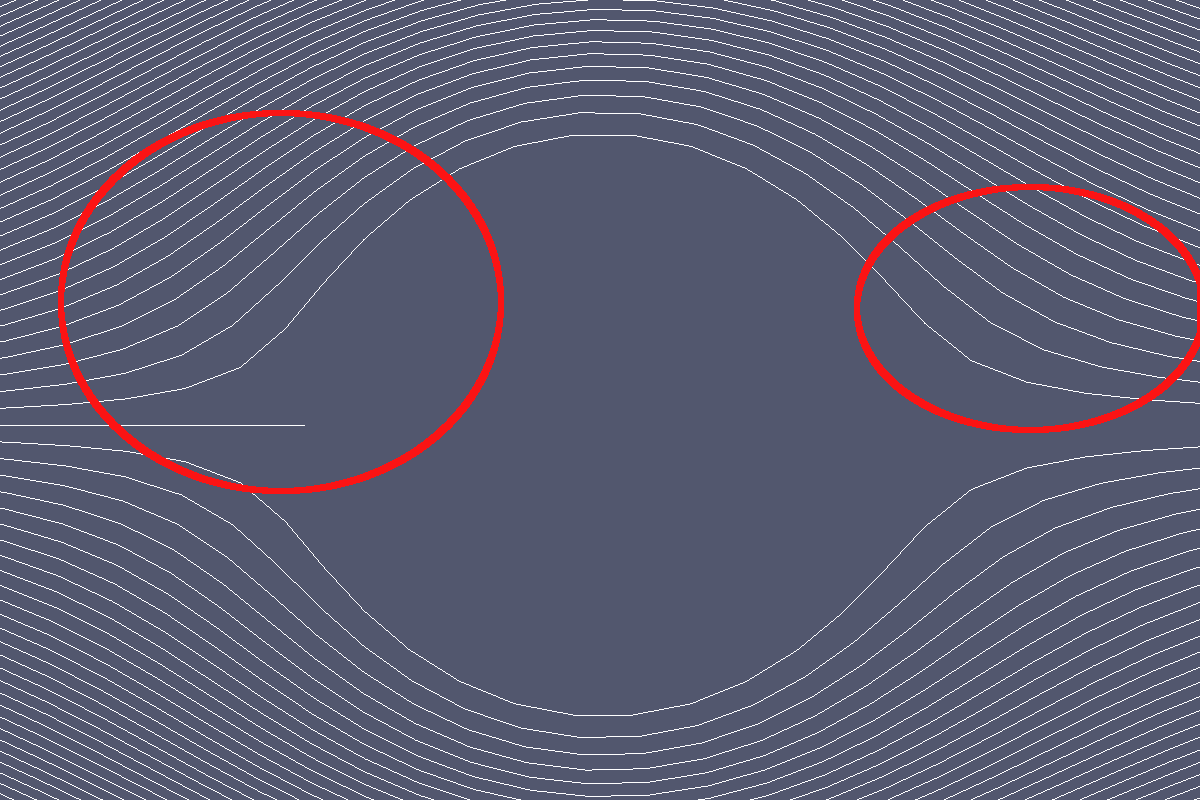}
        \caption{Streamlines for non-Newtonian fluid \#1.}
    \end{subfigure} \\
    \begin{subfigure}[b]{0.49\linewidth}
        \includegraphics[width=\linewidth]{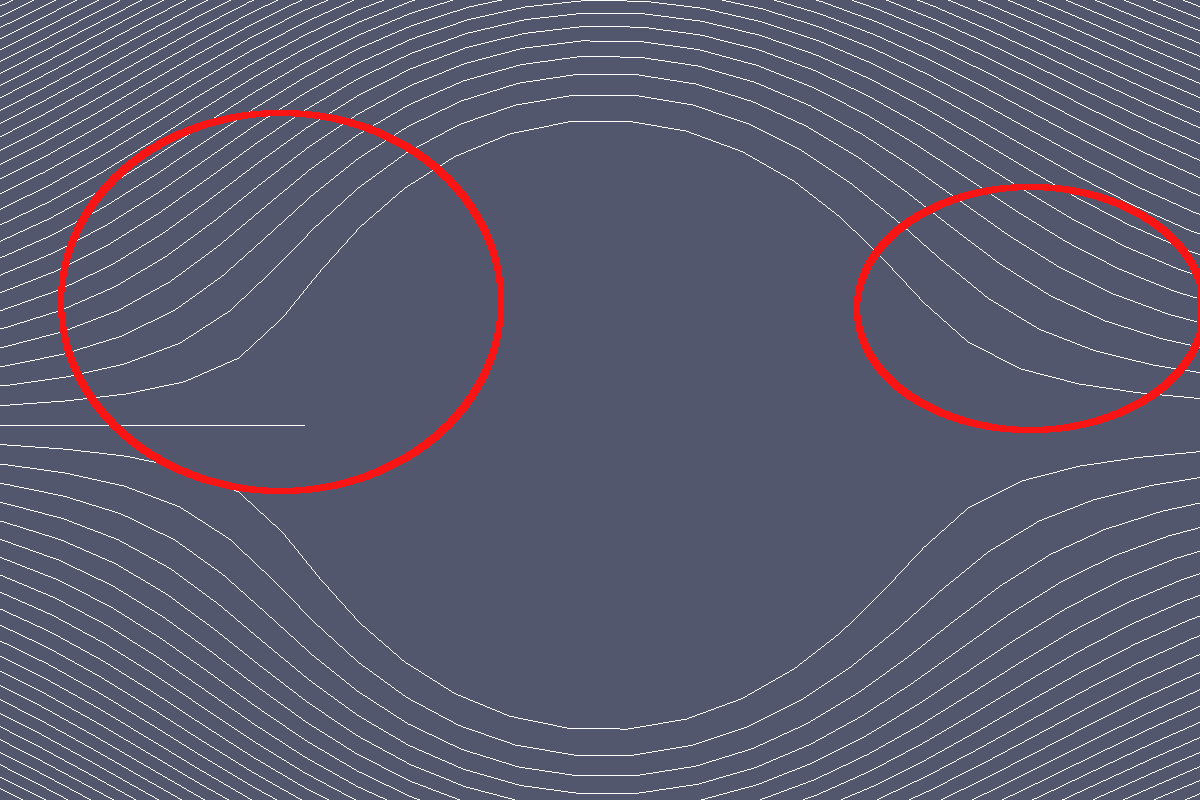}
        \caption{Streamlines for non-Newtonian fluid \#2.}
    \end{subfigure}
    \begin{subfigure}[b]{0.49\linewidth}
        \includegraphics[width=\linewidth]{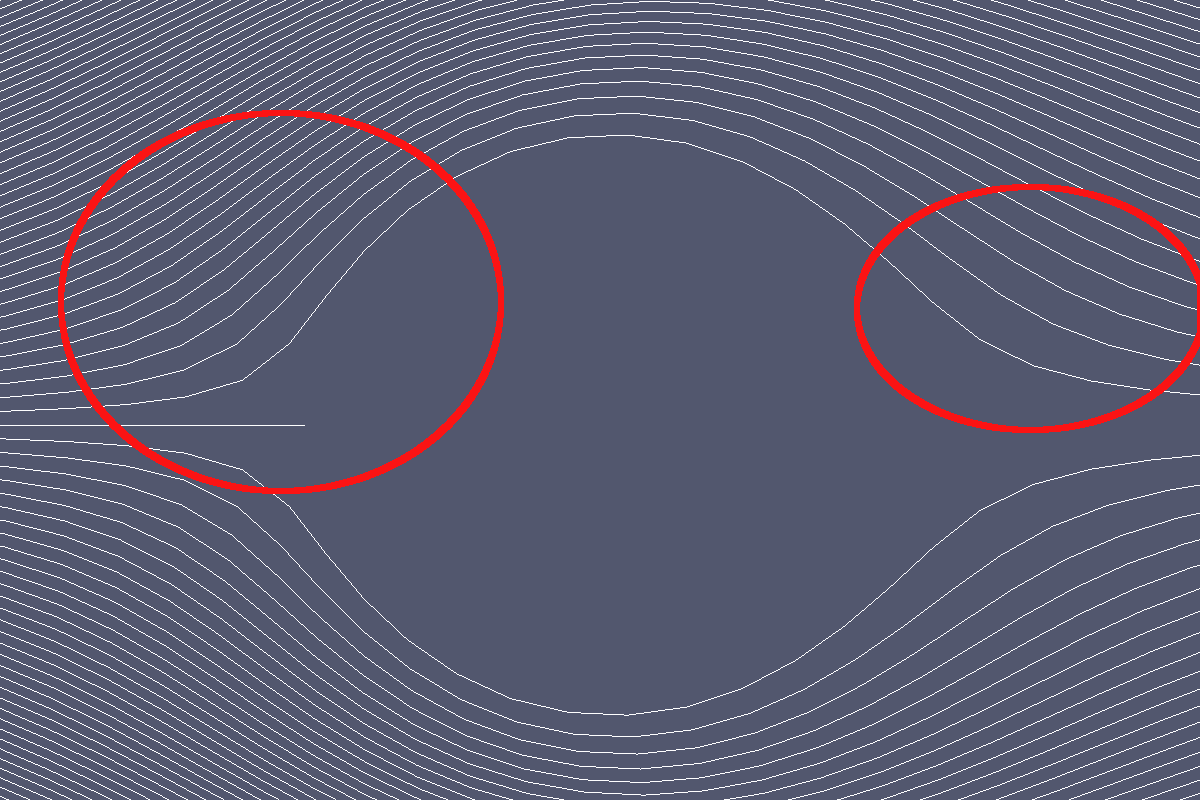}
        \caption{Streamlines for non-Newtonian fluid \#2, higher Re.}
    \end{subfigure}
    \caption{Streamlines for flow around a sphere: (a) Newtonian fluids of differing viscosities and differing laminar flow rates; (b) Cross-model fluid \#1 ($\hat\mu_0=0.00548$ Pa$\cdot$s, $\hat\mu_\infty=0.00222$ Pa$\cdot$s, $m=0.00274$ s, $n=0.676$) flowing with a Re$=4.50\times 10^{-5}$ ; (c)  Cross-model fluid \#2 ($\hat\mu_0=1.35$ Pa$\cdot$s, $\hat\mu_\infty=0.00331$ Pa$\cdot$s, $m=0.322$, $n=0.707$) flowing with a Re$=3.02\times 10^{-4}$ ; and (d)  Cross-model fluid \#2 ($\hat\mu_0=1.35$ Pa$\cdot$s, $\hat\mu_\infty=0.00331$ Pa$\cdot$s, $m=0.322$, $n=0.707$) flowing with a Re$=3.02\times 10^{-2}$.}
    \label{fig:Streamlines}
\end{figure*}

The description of the flow of non-Newtonian fluids through porous media is challenging, because a wide range of shear profiles is typically encountered for such systems. Most macroscopic models used were developed for Newtonian flows. A common model used is Darcy's law, which may be written as
\begin{equation}
\vec q = -\frac{\hat{\kappa}}{\hat{\mu}} (\del p-\rho\vec g)\;,
\end{equation}
where $\vec q$ is the macroscale specific discharge of the fluid, $\hat{\kappa}$ is a permeability assumed to be intrinsic to the media, $\hat \mu$ is the dynamic viscosity of the fluid, $p$ is the fluid pressure, $\rho$ is the density of the fluid, and $\vec g$ is the acceleration due to gravity. Darcy's law was developed to describe laminar, steady-state flow of a Newtonian fluid through an isotropic porous medium  \cite{Darcy_56,Bear_72,Freeze_Cherry_79,Simmons_08}; however, it is often used to describe systems other than those on which it is based. Darcy's law has been routinely used to describe non-Newtonian flows, assuming that the intrinsic permeability is a constant depending only upon the morphology and topology of the pore space and independent of the fluids properties, and then describing the viscosity in Darcy's law as an effective viscosity \cite{Sadowshi_Bird_65,Sorbie_Clifford_etal_89,Chauveteau_82,Parnas_Cohen_87,Hayes_Afacan_etal_96,Liu_Masliyah_99,Perrin_Tardy_etal_06,Tosco_Marchisio_etal_13,Castro_Radilla_17,Berg_Wunnik_17,Zamani_Bondino_etal_17,Castro_19,Zhang_Prodanovic_etal_19}. This is done by collecting experimental or simulation data for a macroscopically one-dimensional horizontal flow and then calculating the effective viscosity from
\begin{equation}
\hat{\mu}_\text{eff} = -\frac{\hat{\kappa}}{q}\frac{\Delta p}{L}\;,
\label{eq:muEff}
\end{equation}
where $\hat{\mu}_\text{eff}$ is the effective, non-Newtonian viscosity and $L$ is the macroscopic length of the system in the primary direction of flow. The problem that arises from this approach is that this effective viscosity is not informative about the state of the system. Much of this confusion arises from a fundamental misunderstanding of the physical underpinning of intrinsic permeability.

The intrinsic permeability is generally considered to be a property of the system that is independent of the fluid flowing through it. This permeability is defined using Darcy's law when a Newtonian fluid is flowing through the system. The physics that underlie this phenomenon can be easily visualized using streamlines, which are the paths that particles would take if they were carried purely by the advection of the fluid through the system. Examples of such streamlines can be seen in Fig. \ref{fig:Streamlines}, which was generated using OpenFOAM for flow around a single sphere. The Reynolds number for each of the flows was calculated using 
\begin{equation}
\text{Re} = \frac{\rho q R}{\Hat{\mu}_\infty}\;,
\end{equation}
where $R$ is the radius of the sphere. In Fig. \ref{fig:Streamlines}(a), a Newtonian fluid flows past a sphere during laminar flow, and the streamlines generated are matched by all Newtonian fluids at all laminar flow rates past this sphere. Because the streamlines are invariant with these changes, it is clear that there is some intrinsic aspect of the system geometry that contributes to the energy loss resulting from fluid flow, independent of the Newtonian fluid used or the flow rate. In the case of non-Newtonian fluid flow, the rheological properties of the fluid may change the observed flow paths. In Fig. \ref{fig:Streamlines}(b), streamlines for one Cross model fluid are shown, which differ only slightly from Fig. \ref{fig:Streamlines}(a). However, when another Cross model fluid flows around the same sphere, as in Fig. \ref{fig:Streamlines}(c), the streamlines differ more markedly. Additionally, when that same Cross model fluid flows around the sphere at a different flow rate, as in Fig. \ref{fig:Streamlines}(d), the streamlines again change. 

In Fig. \ref{fig:Streamlines} the differences in streamlines are more apparent when looking within and near the red ellipses. As the streamlines bend, the fluid molecules are accelerating, leading to energy losses that are observed as a pressure drop for a given flow rate. It is expected that a sharper curve would require greater acceleration of the molecules, and lead to greater energy losses. The differences in how these streamlines occur for different non-Newtonian fluids for a given flow rate is due to the fluids having roughly equivalent inertia before reaching the solid. Once a fluid parcel reaches the solid, the viscosity will differ compared to the bulk solution where the solid has no effect, and the original inertia of the fluid parcel will carry it closer to the solid boundary if the fluid is shear thinning, farther if it is shear thickening. This would lead to a larger acceleration of the fluid parcel as it circumvents the solid, and thus a larger energy loss compared to a fluid that had the same viscosity throughout the bulk solution as the viscosity of the non-Newtonian fluid near the solid boundary. While there is still an impact of the geometry of the system on the flow of the fluid that is apparent, it is not obvious how the fluid properties will change this, or how the new permeability associated with this streamline arrangement can be predicted. 

\section{Theory} 
\subsection{Framework}
\label{sec:framework}

TCAT has been used to formulate a macroscale model hierarchy for single-fluid-phase flow through a porous medium \cite{Gray_Miller_06,Gray_Miller_14}. This general model hierarchy can be used to formulate closed models of varying sophistication for a wide range of systems, including non-isothermal systems, systems with complex solid behavior, both laminar and turbulent flows, and both Newtonian and non-Newtonian flows. While the framework is in place for the development of such models, work to date has considered only relatively simple systems consisting of the laminar flow of Newtonian fluids through systems with simple solid properties. 

The framework consists of a simplified entropy inequality (SEI) that relates the sum of fluxes and forces for dissipative processes to the entropy production rate of the system, and a full set of macroscale conservation of mass, momentum, and energy equations. The formulation also includes precise descriptions of all variables in terms of averages of microscale quantities, including expressions for the inter-entity exchange of conserved quantities.  This explicit connection between the microscale and the macroscale provides a means to evaluate model approximations and to use averaged microscale simulation, or experimental, results to evaluate and validate a TCAT model. TCAT models are guided by closure approximations that are consistent with the second law of thermodynamics. 

We can use this TCAT model framework to consider the case of non-Newtonian flow of a single fluid through porous media. We will assume the system is isothermal, that the chemical composition of each entity is constant in space and time, mass transfer does not occur between phases, the interface between the water phase and solid is massless, flow is laminar, and the solid phase is incompressible and immobile; these conditions form a set of secondary restrictions that can be used to deduce a specific model instance from a general model hierarchy. Thus, our focus is to leverage an existing TCAT model hierarchy to develop a relatively simple Non-Newtonian flow model based upon upscaling of microscale conservation and thermodynamic principles. This is a departure from extant phenomenological approaches that posit forms directly at the macroscale \cite{Hayes_Afacan_etal_96,Liu_Masliyah_99}. To formulate this model, we need to define the averaging process used, the conservation equations to be closed, the entropy inequality used to ensure consistency with the second law of thermodynamics, and the closure relations needed to yield a closed, solvable model. 

\subsection{Averaging}

Because the TCAT model includes averaged microscale quantities computed in different ways, some definitions are needed for clarity. For some quantity $f$ that is averaged over the domain $\Dm \gkb$, normalized by an integral over the domain $\Dm \gkg$, and weighted by $W$, the averaging operator is defined as \cite{Miller_Gray_05,Gray_Miller_14}
\begin{equation}
\big\langle{f}\big\rangle_{\Dm\gkb,\Dm\gkg, W} = \gaint f{\Dm \gkb}{\Dm \gkg}{W}\;.
\label{eq:aop}
\end{equation}

Various instances of \Eqn{aop} occur routinely, making it convenient to define a set of commonly occurring averaged forms. One common average is
\begin{equation}
    f^\gkb_\gka = \big\langle f_\gka\big\rangle_{\Dm \gkb,\Dm \gkb}\;,
\end{equation}
where the subscript on $f$ denotes the microscale entity (phase, interface, or common curve), the superscript denotes the entity over which averaging is performed, which is typically of one dimension lower than $\Dm \qe$, and the abscence of $W$ implies a unit value. 

An intrinsic average is defined as
\begin{equation}
f^\gka = \big\langle f_\gka\big\rangle_{\Dm \gka,\Dm \gka}\;,
\end{equation}
and a density-weighted average is defined as
\begin{equation}
f^{\sol{\gka}} = \big\langle f_\gka\big \rangle_{\Dm\gka,\Dm\gka,\rho_\gka}\;,
\end{equation}
where $\rae$ is a mass density.

During averaging, some variables arise that do not fit one of the above special forms. Such macroscale averages are denoted with a double overbar and each occurrence is explicitly defined. A specific entity measure is such a quantity, and it is defined as
\begin{equation}
\ep = \big\langle 1\big\rangle_{\Dm \gka,\Dm{}}\;,
\end{equation}
where $\Dm{}$ is an averaging region that contains all entities. 
Specific entity measures are used to convert averages over the entire domain to intrinsic averages. An example of such an operation is shown below for the quantity $f_\gka$,
\begin{equation}
    \big\langle f_\gka\big\rangle_{\gkO_\gka,\gkO} = \big\langle1\big \rangle_{\gkO_\gka,\gkO}\big <f_\gka\big >_{\gkO_\gka,\gkO_\gka} = \ep f^\gka\;.
\end{equation}
 
\subsection{Conservation Equations}

Mechanistic models of flow through porous media rely upon a set of conservation equations, and a set of closure relations to render the equations solvable. The relevant conservation equations include a conservation of mass equation for the water phase \cite{Gray_Miller_14}, which can be written as
\begin{equation}
\pdt{\lrp{\ew\raw}} 
+ \del\vdot\lrp{\ew\raw\vaw} = 0\;,
\label{eq:M}
\end{equation}
and a conservation of momentum equation for the water phase of the form
\begin{equation}
\pdt{\lrp{\ew\raw\vaw}} 
+ \del\vdot\lrp{\ew\raw\vaw\vaw} 
- \del\vdot\lrp{\ew\taw}
- \ew\raw\gaw 
-\ieT{s}{w} 
 = \vec 0\;,
\label{eq:P}
\end{equation}
where $\ew$ is the volume fraction, $t$ is time, $\vaw$ is the velocity vector, $\taw$ is the stress tensor, $\gaw$ is the gravitational acceleration vector, $\ieT sw$ is the rate of momentum density transfer from the solid phase to the water phase, $w$ denotes the water phase, $s$ denotes the solid phase, and superscripts denote averaged macroscale variables.  Producing a solvable model requires a set of closure approximations for $\raw$, $\taw$, and $\ieT sw$, which can be deduced using thermodynamics, and the  available entropy inequality \cite{Gray_Miller_06,Gray_Miller_14}. 

\subsection{Entropy Inequality}

The SEI for a single-fluid phase in a solid porous medium has been developed and is provided in \cite{Gray_Miller_06,Gray_Miller_14}. This expression is also a subset of more complicated and available SEI's, such as single-fluid flow and transport, two-fluid flow in porous media, two-fluid flow and species transport in porous media, and flow in general three-phase systems \cite{Gray_Miller_09,Jackson_Miller_etal_09,Gray_Miller_14,Rybak_Gray_etal_15,Miller_Valdes-Parada_etal_17,Miller_Gray_etal_19}.

The standard TCAT approach for deriving an SEI is to derive the most general form possible for a general class of model to enable the use of this expression for any subset of a general model hierarchy; other approaches are possible \cite{Miller_Valdes-Parada_etal_17}. Because general SEI expressions for the entropy production rate density are long, complicated equations, a useful strategy is to consider simple subsets of the most general case. Such a restricted SEI can be derived by specifying a set of secondary restrictions and using these to reduce a general SEI to the minimum form needed for a specific model instance. Applying the restrictions noted in \S~\ref{sec:framework}, yields a restricted SEI of the form
\begin{equation}
\frac{1}{\gkt} \lrp{\ew\taw + \ew\paw\ten I}\dd \daw + \frac{1}{\gkt} \bigg [ \ew\del\paw - \ew\raw\del\lrp{\cpaw + \gpaw} 
- \ew\raw\gaw + \ieT{w}{s} \bigg ]\cdot\vaw = \gkL \geq 0\;,
\label{eq:SEI}
\end{equation}
where $\gkt$ is the temperature, $\paw$ is the pressure, $\tI$ is the identity tensor, $\daw$ is the rate of strain tensor, $\cpma w$ is the chemical potential, $\gpaw$ is the gravitational potential, and $\gkL$ is the entropy production rate density of the system.

This entropy inequality is in flux-force form, and it provides permissibility constraints for closure relations. The specific form of closure relations is not unique, but any valid condition must not violate \Eqn{SEI}. All members of the set of fluxes are unique and all members of the set of forces are also unique. These properties allow fluxes to be considered one at a time and closure relations posited that are either conjugate flux-force or cross-coupled flux-forces in form \cite{Gray_Miller_14}. The usual approach is to generate the simplest possible form that yields a valid model for the application of concern.

The flux in the first term on the left-hand-side of \Eqn{SEI} involves the stress tensor and the fluid pressure, and the conjugate force is the rate of strain tensor. Both the flux and the force vanish at least at equilibrium.   The simplest possible closure relation consistent with entropy inequality is a zero-order closure in which the flux is assumed to be equal to zero for all cases, which implies 
\begin{equation} 
\taw = -\paw\ten I\;.
\label{eq:t}
\end{equation}
This closure approximation is a statement that the flow is inviscid at the macroscale. This is reasonable if momentum transfer between the fluid and solid phase dominates over the interaction of the fluid with a boundary of the system. Given the typically large interfacial area between the fluid and solid phase compared to the interfacial area between the fluid and the boundary of most domains of interest, this closure approximation is not only simple but also well founded. 

The second term on the left-hand-side of \Eqn {SEI} describes the entropy production due to the flow of the fluid through the medium, and its interaction with the solid phase. We wish to derive an approximation for the flux, which is the term in brackets. The simplest possible closure relation consistent with the second law of thermodynamics and a known production of entropy resulting from flow is a first-order conjugate flux-force closure of the form
\begin{equation} 
\ew\del\paw - \ew\raw\del\lrp{\cpaw + \gpaw} - \ew\raw\gaw + \ieT{w}{s} = \htR^w\vdot\vaw\;,
\label{eq:T}
\end{equation}
where $\htR^w$ is a second rank positive semi-definite tensor. This form ensures that any flow generates entropy; equilibrium requires that the fluid velocity vanishes. The axiom of objectivity in continuum mechanics requires all velocities to be relative velocities, but in this case the relative velocity is assumed to be the solid-phase velocity, which is zero in the reference coordinate system. 

\Eqnstwo{t}{T} can be used to write \Eqn{P} as
\begin{equation}
\pdt{\lrp{\ew\raw\vaw}} 
+ \del\vdot\lrp{\ew\raw\vaw\vaw} 
+\ew\raw\del\lrp{\cpma w+\gpma w}
+\htR^w\vdot\vaw
 = \vec 0\;.
\label{eq:P1}
\end{equation}
To calculate the change in chemical potential $(\cpaw)$, the macroscale Gibbs-Duhem equation can be used, which can be written for isothermal conditions as \cite{Gray_Miller_14}
\begin{equation}
- \ew\dif\paw +\ew\raw\dif\cpaw 
- \big\langle\dif\lrp{\piw-\paw}\big\rangle_{\Dm w,\Dm{}}  
+ \big\langle\riw\dif\lrp{\cpiw-\cpaw}\big\rangle_{\Dm w,\Dm{}} 
= 0\;.
\label{eq:gd}
\end{equation}
Because of the restrictions imposed on the solid phase, the two averaged deviation terms in \Eqn{gd} vanish giving
\begin{equation}
-\ew\dif\paw +\ew\raw\dif\cpaw = 0\;,
\label{eq:gdh}
\end{equation}
or
\begin{equation}
-\ew\del\paw + \ew\raw\del\cpaw = 0\;.
\label{eq:gdf}
\end{equation}
Substituting \Eqn{gdf} into \Eqn {P1} yields
\begin{equation}
\pdt{\lrp{\ew\raw\vaw}} 
+ \del\vdot\lrp{\ew\raw\vaw\vaw} 
+\ew\del\paw
+\ew\raw\del\gpma w
+\htR^w\vdot\vaw
 = \vec 0\;.
\label{eq:P2}
\end{equation}
The restriction on the solid phase behavior allows the gradient of the gravitational potential to be written as the gravitational acceleration giving the momentum equation
\begin{equation}
\pdt{\lrp{\ew\raw\vaw}} 
+ \del\vdot\lrp{\ew\raw\vaw\vaw} 
+\ew\del\paw
-\ew\raw\gaw
+\htR^w\vdot\vaw
 = \vec 0\;.
\label{eq:P3}
\end{equation}

Substituting \Eqn{t} into \Eqn{P} yields a conservation of momentum equation of the form
\begin{equation}
\pdt{\lrp{\ew\raw\vaw}} 
+ \del\vdot\lrp{\ew\raw\vaw\vaw} 
+ \del\lrp{\ew\paw}
- \ew\raw\gaw 
-\ieT{s}{w} 
 = \vec 0\;,
\label{eq:P4}
\end{equation}
Subtracting \Eqn{P4} from \Eqn{P3}, applying the product rule, and rearranging yields
\begin{equation}
\ieT sw=
-\htR^w\vdot\vaw
-\paw\del\ew\;,
\label{eq:TR}    
\end{equation}
or for the solid phase restrictions previously specified
\begin{equation}
\ieT ws=
\htR^w\vdot\vaw\;,
\label{eq:TR1}    
\end{equation}
where $\ieT ws=-\ieT sw$ has been used.

\subsection{Momentum Transfer Analysis}

\Eqn{TR1} expresses the equality between the transfer of momentum from the fluid to the solid and the first-order closure relation derived from the SEI and given by \Eqn{T}. In light of this equality, and the conservation of momentum equation, an analysis can be performed for both Newtonian and non-Newtonian flow through a porous media. For slow, essentially steady flow through a porous medium---and considering the magnitude of each term, \Eqn{P3} can be written as
\begin{equation}
\ew\del\paw
-\ew\raw\gaw
+\htR^w\vdot\vaw
 = \vec 0\;.
\label{eq:P5}
\end{equation}
Darcy's law is an expression written for the product $\ew\vaw$, which is the specific discharge and is also referred to as the Darcy velocity. Rearranging \Eqn{P5}, the specific discharge may be written as
\begin{equation}
\ew\vaw=
-\lrp{\ew}^2\lrp{\htR^w}^{-1}
\vdot\lrp{\del\paw-\raw\gaw}\;,
\label{eq:Darcy}    
\end{equation}
which can be written in scalar form as
\begin{equation}
\ew\sbs vw=
-\lrp{\ew}^2\lrp{\hsR^w}^{-1}
\lrp{\od{\paw} x+\raw g}\;,
\label{eq:sDR1d}    
\end{equation}
where the gravitational vector has been assumed to be aligned with the negative $x$ coordinate direction and independent of the phase.

Darcy's law may be written in scalar form as
\begin{equation}
\ew \sbs vw = 
-\frac{\hsK^w}{\raw g}\lrp{\od \paw x + \raw g}\;,
\label{eq:sD1d}
\end{equation}
where $\hsK^w$ is the hydraulic conductivity, which is typically defined as
\begin{equation}
\hsK^w=\frac {\hat\gkk \raw g}{\hat\mu^w}\;,
\label{eq:Kdef}
\end{equation}
where $\hat\gkk$ is the so-called intrinsic permeability. It is commonly reported that $\hat\gkk$ describes the pore morphology and topology, while $\raw$ and $\hat\mu^w$ describe the fluid properties \cite{Hubbert_56,Sadowshi_Bird_65,Bear_72,Freeze_Cherry_79,Sorbie_Clifford_etal_89,Chauveteau_82,Parnas_Cohen_87,Hayes_Afacan_etal_96,Liu_Masliyah_99,Bird_Stewart_etal_02,Perrin_Tardy_etal_06,Tosco_Marchisio_etal_13,Castro_Radilla_17,Berg_Wunnik_17,Zamani_Bondino_etal_17,Castro_19,Zhang_Prodanovic_etal_19}. \Eqns{sDR1d} {Kdef} can be used to deduce
\begin{equation}
\hsR^w=
\frac{\lrp{\ew}^2\hat\mu^w}{\hat\kappa}\;.
\label{eq:RK}
\end{equation}

Similarly, a common tensor form of Darcy's law can be written as
\begin{equation}
\ew\vaw=
-\frac{\htk^w}{\hat\mu^w}\vdot\lrp{\del\paw-\raw\gaw}\;,
\label{eq:kdt}    
\end{equation}
and using \Eqn{Darcy} it follows that
\begin{equation}
\htk^w=\lrp{\ew}^{2}\hat\mu^w\lrp{\htR^w}^{-1}\;.
\label{eq:ktR}
\end{equation}

Because TCAT precisely defines all quantities in terms of averages of microscale quantities, the definition for the momentum transfer exists and may be written as \cite{Gray_Miller_14}
\begin{equation}
\ieT{w}{s} = -\big\langle\lrb{\tiw + \riw\lrp{\viw - \vec v^{\sol{ws}}_{w}}\lrp{\viws - \viw} }\vdot\vnmi w \big\rangle_{\Dm\ws,\Dm{}}\;,
\label{eq:Tadef}
\end{equation}
where because of the massless interface condition, a jump condition for momentum transfer from the water to the solid phase has been written rather than momentum transfer to the $\ws$ interface, and $\Dm\ws$ is the domain of the interface between the water and solid phases. The velocity product term on the righ-hand-side of \Eqn{Tadef} is related to mass transfer, which does not occur in this system, allowing \Eqn{Tadef} to be written as
\begin{equation}
\ieT{w}{s} = -\big\langle\tiw\vdot\vnmi w \big\rangle_{\Dm\ws,\Dm{}}\;.
\label{eq:Ta1}
\end{equation}

The microscale stress tensor for an isotropic fluid is \cite{Gray_Miller_14}
\begin{equation}
\tiw = -\piw\ten I + 2\Hat{\mu}_w\diw - \frac 23 \hat{\mu}_w\lrp{\ten I\dd\diw}\ten I\;.
\label{eq:tidef}
\end{equation}
Because water is only slightly compressible, the last term in \Eqn{tidef} can be dropped and the remainder of this equation can be substituted into \Eqn{Ta1} yielding
\begin{equation}
\ieT{w}{s} = 
-\big\langle\lrp{-\piw\ten I 
+ 2\hat{\mu}_w\diw }\vdot\vnmi w \big\rangle_{\Dm\ws,\Dm{}}\;,
\end{equation}
which can be written as
\begin{equation}
\ieT{w}{s} = 
\ews\big\langle\piw\vnmi w\big\rangle_{\Dm\ws,\Dm\ws} 
-\ews\big\langle 2\hat{\mu}_w\diw\vdot\vnmi w\big\rangle_{\Dm\ws,\Dm\ws}\;,
\label{eq:Ta2}
\end{equation}
where $\ews$ is the specific interfacial area of the $\ws$ interface.

Both $\piw$ and $\hat\mu_w$ in a non-Newtonian fluid will vary for a flowing system at the microscale. These terms can be expanded into a mean and a fluctuation term, and \Eqn{Ta2} can be written as
\begin{equation}
\ieT{w}{s} = 
\ews p_w^\ws\big\langle\vnmi w\big\rangle_{\Dm\ws,\Dm\ws} 
+\ews \big\langle\lrp{\piw-p_w^\ws}\vnmi w \big\rangle_{\Dm\ws,\Dm\ws} 
-\ews\hat\mu_w^\ws\big\langle 2\diw\vdot\vnmi w\big\rangle_{\Dm\ws,\Dm\ws}
-\ews\big\langle 2\lrp{\hat\mu_w-\hat\mu_w^\ws}\diw\vdot\vnmi w\big\rangle_{\Dm\ws,\Dm\ws}\;.
\label{eq:Ta3}
\end{equation}
Averaging theorems can be used to show \cite{Gray_Miller_14}
\begin{equation}
\del\ew=
-\big\langle\mnw\big\rangle_{\Dm\ws,\Dm{}}\;,
\label{eq:ew-nw}
\end{equation}
which since the porosity is constant in this case, \Eqn{Ta3} may be simplified to
\begin{equation}
\ieT{w}{s} = 
\ews \big\langle\lrp{\piw-p_w^\ws}\vnmi w \big\rangle_{\Dm\ws,\Dm\ws} 
-\ews\hat\mu_w^\ws\big\langle 2\diw\vdot\vnmi w\big\rangle_{\Dm\ws,\Dm\ws}
-\ews\big\langle 2\lrp{\hat\mu_w-\hat\mu_w^\ws}\diw\vdot\vnmi w\big\rangle_{\Dm\ws,\Dm\ws}\;.
\label{eq:Ta4}
\end{equation}
This expression includes a mean viscous term, a pressure fluctuation term, and a viscous fluctuation term. We posit that the mean viscous term is the dominant term in this equation. For a homogeneous, isotropic medium the pressure fluctuation term will be dependent on the gradient in pressure, thus we posit
\begin{equation}
\ews \big\langle\lrp{\piw-p_w^\ws}\vnmi w \big\rangle_{\Dm\ws,\Dm\ws} \propto
\ell\ews\del\paw\;,
\label{eq:pf}
\end{equation}
where $\ell$ is a characteristic length.

The microscale shear rate tensor is defined as
\begin{equation}
\dot{\ten\gkg}_w = 2\diw\;,
\label{eq:dgi}
\end{equation}
and the corresponding macroscale shear rate tensor is defined as
\begin{equation}
\dot{\ten\gkg}^{\dol{\ws}}_w = \big\langle\del\viw + \lrp{\del\viw}\T\big\rangle_{\Dm\ws,\Dm\ws}\;.
\label{eq:dga}
\end{equation}
The magnitude of the shear rate tensor used in non-Newtonian models, such as the Cross model, for the dynamic viscosity is at the microscale
\begin{equation}
\dot\gkg_w=
\sqrt{\dot{\ten\gkg}_w\dd \dot{\ten\gkg}_w}\;,
\label{eq:dgkgi}
\end{equation}
and similarly at the macroscale
\begin{equation}
\dot\gkg_w^\ws=
\sqrt{\dot{\ten\gkg}_w^\ws\dd \dot{\ten\gkg}_w^\ws}\;.
\label{eq:dgkgi}
\end{equation}
Due to the no-slip condition at the $ws$ interface, the gradient in velocity only occurs in the $\vnmi w$ direction, thus 
\begin{equation}
 \dot\gkg_w=\sqrt{\dot{\ten\gkg}_w\dd \dot{\ten\gkg}_w} = |2\diw\cdot\vnmi w|
\label{eq:srmi}
\end{equation} 
and noting
\begin{equation}
2\diw\vdot\mnw=\dot\gkg_w \vec e_{dn}\;,
\label{eq:dwn}
\end{equation}
where $\vec e_{dn}$ is a unit direction vector corresponding to the direction of $2\diw\vdot\mnw$. Using \Eqnstwo{srmi}{dwn} and defining $\vec e^{\dol{ws}}_{dn}$ as 
\begin{equation}
\vec e^{\dol{ws}}_{dn}=\big\langle \vec e_{dn} \big\rangle_{\Dm\ws,\Dm\ws,\dot\gkg_w}
\label{eq:ednws}
\end{equation}
the average of the unit direction vector of $2\diw\cdot\vnmi w$ allows \Eqn{Ta4} to be written as
\begin{equation}
\ieT{w}{s} = 
\ews \big\langle\lrp{\piw-p_w^\ws}\vnmi w \big\rangle_{\Dm\ws,\Dm\ws} 
-\ews\Hat{\mu}_w^{ws}\dot{\gkg}^{\dol{ws}}_w\vec e^{\dol{ws}}_{dn}
-\ews\big\langle 2\lrp{\hat\mu_w-\hat\mu_w^\ws}\diw\vdot\vnmi w\big\rangle_{\Dm\ws,\Dm\ws}\;.
\label{eq:Ta5}
\end{equation}
If the microscale deviation terms are negligible, the momentum transfer term becomes
\begin{equation}
\ieT{w}{s} = -\ews\hat{\mu}_w^{ws}\dot{\gkg}^{\dol{ws}}_w \vec e^{\dol{ws}}_{dn}\;,
\label{eq:Ta6}
\end{equation}
which, using \Eqn{TR1}, yields
\begin{equation}
-\ews\hat{\mu}_w^{ws}\dot{\gkg}^{\dol{ws}}_w\vec e^{\dol{ws}}_{dn} = \hat{\ten{R}}^w\vdot\vaw\;.
\label{eq:TR2}
\end{equation}

We will next decompose the macroscale velocity vector into a magnitude $v^{\dol{w}} = |\vaw|$ and unit direction vector $\vec e^{\dol{w}}_{v}$. Additionally, the resistance tensor is symmetric semi-positive definite, thus it is possible to carry out an eigenvalue decomposition such that
\begin{equation}
\hat{\ten{R}}^w = \ten Q^w\vdot\ten \gkL^w\vdot \lrp{\ten Q^w}^{-1}\;,
\label{eq:Red}
\end{equation}
where the column vectors of $\ten Q^w$ are the eigenvectors of the resistance tensor, and $\ten \gkL^w$ is a diagonal tensor with the corresponding eigenvalues \cite{Dye_McClure_etal_13}. Assuming that the principal axes have been selected such that they correspond to the principal directions of the resistance tensor, making $\hat{\ten{R}}^w=\ten \gkL^w$, \Eqn{TR2} becomes
\begin{equation}
-\ews\hat{\mu}_w^{ws}\dot{\gkg}^{\dol{ws}}_w\vec e^{\dol{ws}}_{dn} = v^\dol{w}\hat{\ten{R}}^w\vdot\vec e^{\dol{w}}_{v}\;.
\label{eq:Rvd}
\end{equation}

Dotting both sides of \Eqn{Rvd} with the inverse of the resistance tensor and performing algebraic rearrangement yields
\begin{equation}
\lrp{\hat{\ten{R}}^w}^{-1}\vdot\vec e^{\dol{ws}}_{dn} = \lrp{\ten \gkL^w}^{-1}\vdot\vec e^{\dol{ws}}_{dn} =  -\frac{v^{\dol{w}}}{\ews\hat{\mu}_w^{ws}\dot{\gkg}^{\dol{ws}}_w} \vec e^{\dol{w}}_{v}\;.
\label{eq:Rvd1}
\end{equation}

Because we have assumed alignment with the principal directions of anistropy, and resultingly $\htR^w$ is diagonal, it follows that 
\begin{equation}
\lrp{\hat{\ten{R}}^w}^{-1} = \lrp{\ten \gkL^w}^{-1} =  -\frac{v^{\dol{w}}}{\ews\hat{\mu}_w^{ws}\dot{\gkg}^{\dol{ws}}_w} \ten e_{\ten K}\;,
\label{eq:Rek}
\end{equation}
where $\ten e_{\ten K}$ is a diagonal tensor that differs from $(\htR^w)^{-1}$ by a linear scaling factor, where the diagonal entries of this tensor are 
\begin{equation}
 e_{\ten Kii} = \frac{e^{\dol{w}}_{vi}}{e^{\dol{ws}}_{dni}}\;,
 \label{eq:eK}
\end{equation}
where $i$ denotes a vector and tensor component index.

\Eqn{Rek} can be viewed in light of the traditional forms given in \Eqns{Kdef}{ktR}, which provides a definition for the hydraulic conductivity tensor given by 
\begin{equation} 
\htK^w = \lrp{\ew}^2\lrp{\htR^w}^{-1}\raw g = -\frac{\lrp{\ew}^2 \raw g  v^{\dol{w}}} {\ews\hat{\mu}_w^{ws}\dot{\gkg}^{\dol{ws}}_w} \ten e_{\ten K}\;.
\label{eq:KtR2}
\end{equation}
\Eqnstwo{ktR}{Rek} can be combined to define the intrinsic permeability tensor in terms of the precise averages derived above
\begin{equation} 
\htk^w = -\frac{\lrp{\ew}^2  \hat\mu^w v^{\dol{w}}} {\ews\hat{\mu}_w^{ws}\dot{\gkg}^{\dol{ws}}_w} \ten e_{\ten K}\;,
\label{eq:ktR2}
\end{equation}
which in the classical literature is assumed to be symmetric positive semi-definite second-order tensor that is only dependent upon the pore morphology and topology, where the restriction of Darcy flow to the creeping flow regime is commonplace \cite{Bear_72,deMarsily_86,Dye_McClure_etal_13}. Some observations can be made regarding \Eqn{ktR2}. For Newtonian fluids the dynamic viscosity is constant, which makes the averaging method unimportant and the ratio of these two averages is unity and drops out of the equation. Approximations for $\hsk^w$ are often posed in terms of the Sauter mean diameter, which appears naturally in this equation and is consistent with the success of such parameterizations\cite{Pan_Hilpert_etal_01,Dye_McClure_etal_13}.  Lastly, for Darcy flow of a Newtonian fluid, the ratio of the magnitude of the velocity and the shear rate are constant, which is consistent with $\hsk^w$ being constant for Darcy flow of a Newtonian fluid. 

\subsection{Non-Newtonian Resistance}

A significant insight that arises from momentum transfer analysis is that the shear stress and dynamic viscosity at the solid-fluid interface may be calculated from the pressure drop through the system without knowing what the flow rate is for that given pressure drop. To do this, rearrange \Eqn{P5} such that the term $\hat{\ten R}^w\cdot\vaw$ is isolated, then substitute \Eqn{TR2} to get the equation
\begin{equation}
\ews\hat{\mu}_w^{ws}\dot{\gkg}^{\dol{ws}}_w \vec e^{\dol{ws}}_{dn} = \ew\del\paw -\ew\raw\gaw\;.
\label{eq:nndp}
\end{equation}
Decomposing the forces $\lrp{\del\paw - \raw\gaw}$ into a magnitude and force direction vector $\vec e^\dol{w}_\psi$ gives
\begin{equation}
\ews\hat{\mu}_w^{ws}\dot{\gkg}^{\dol{ws}}_w \vec e^{\dol{ws}}_{dn} = \ew|\del\paw -\raw\gaw|\vec e^\dol{w}_\psi\;.
\label{eq:nndp1}
\end{equation}
While it is possible that neither of the unit vectors present in \Eqn{nndp1} may be known, it must be the case that they are equal to one another. Taking the dot product of a unit vector by itself must be equal to 1, thus
\begin{equation}
\hat{\mu}_w^{ws}\dot{\gkg}^{\dol{ws}}_w = \frac{\ew}{\ews}|\lrp{\del\paw - \raw\gaw}|\;.
\label{eq:nndp2}
\end{equation}
\Eqn{nndp2} can be used to calculate the macroscale surface average of the dynamic viscosity and shear rate at the fluid-solid interface, as long as the orientations of the flow and forces are known. This is done by inserting the shear rate relationship for the dynamic viscosity and then calculating the shear rate for the pressure drop that is anticipated. This is significant, because this relationship can be used to find the non-Newtonian viscosity region that will result from a given pressure gradient. In the case that the pressure fluctuation term in \Eqn{Ta4} is non-negligible, an approximation of this term would need to be included in the analysis leading to \Eqnstwo{TR2}{nndp2}. 

\Eqn{nndp2} can be used to compute $\hsR^w$ for non-Newtonian fluids. An isotropic system will be treated first and then a tensor form of the model will be provided. Taking \Eqn{TR2}, dividing by $\lrp{\ew}^2$ and rearranging gives
\begin{equation} 
\hsR^w q^\dol{w} = -\ew\ews \hat{\mu}_w^{ws}\dot{\gkg}^{\dol{ws}}_w\;,
\label{eq:R_Mu}
\end{equation}
where $q^\dol{w} = \ew v^\dol{w}$ is the magnitude of the Darcy velocity, and $\hsR^w$ is the scalar resistance. Since it is known that $\hsR^w$ is not a constant \cite{Sorbie_Clifford_etal_89,Hayes_Afacan_etal_96,Zamani_Bondino_etal_17}, even though we know each term on the RHS of \Eqn{R_Mu}, we only know the product of $\hsR^w q^{\dol w}$ as a function of $\dot\gkg^{\dol\ws}_w$. Instead, let us make the {\em ansatz} that because $q^\dol{w}$ and $\dot{\gkg}^{\dol{ws}}_w$ are both expected to be monotonically increasing with the change in pressure in the case of a shear thinning fluid (decreasing for a shear thickening one), the resistance $\hsR^w$ and the macroscale dynamic viscosity $\mu_w^\ws$ have a similar functional forms. This would mean, in the case of a Cross model fluid, that 
\begin{equation} 
\hat{R}^w = \hat{R}^w_\infty + \frac{\hat{R}^w_0 - \hat{R}^w_\infty}{1 + \lrp{Mq^\dol{w}}^N}\;,
\label{eq:RES_FORM}
\end{equation}
where $\hsR^w_\infty$ is the infinite flow resistance, $\hsR^w_0$ is the zero-flow resistance, and $M$ and $N$ are system parameters.  

Due to the Newtonian viscosity plateaus present for a Cross model fluid, it is expected that there will be some high shear rate and some low shear rate at which the flow through the porous medium appears strictly Newtonian, thus we can assume, based on \Eqn{RK}, that
\begin{gather}
\hsR^w_\infty = \frac{\lrp{\ew}^2\hat{\mu}_\infty}{\hat{\gkk}_N} \;,
\hsR^w_0 = \frac{\lrp{\ew}^2\hat{\mu}_0}{\hat{\gkk}_N}\;,
\label{eq:RI0}
\end{gather}
where $\hat{\gkk}_N$ is the permeability of the medium when a Newtonian fluid is flowing through it, and $\hat{\mu}_\infty$ and $\hat{\mu}_0$ are Newtonian plateau viscosities described in \Eqn{Cross_Model}, which are constant at the respective infinite- and zero-flow limits. 

To calculate $M$ and $N$, we must apply some further assumptions. It will be assumed here that the resistance model will have the same inflection point as the inflection that occurs for $\hat{\mu}^\ws_w$, and that $N = n$ where $n$ is the exponent in \Eqn{Cross_Model}. For a Cross model fluid, the inflection point in its viscosity $\hat{\mu}_I$ occurs when 
\begin{equation}
\hat{\mu}_I = \sqrt{\hat{\mu}_0\hat{\mu}_\infty}\;.
\label{eq:muI}
\end{equation}
It will also be the case that, if the inflection points of the hydraulic resistance and the viscosity are the same, then when the inflection occurs 
\begin{equation}
1 + \lrp{M q_I}^N = 1 + \lrp{m \dot{\gkg}_I}^n\;,
\label{eq:Nn}
\end{equation}
where $q_I$ and $\dot{\gkg}_I$ are the Darcy velocity and shear rate at which the inflection point occurs. Assuming that $N = n$, it follows that
\begin{equation}
M = \frac{m\dot{\gkg}_I}{q_I}\;.
\label{eq:M}
\end{equation}

It follows from \Eqn{RI0} that for the Cross-like resistance model the inflection resistance is
\begin{equation}
\hsR^w_I = \frac{\lrp{\ew}^2\sqrt{\hat{\mu}_0\hat{\mu}_\infty}}{\hat\gkk_N}\;.
\label{eq:RI}
\end{equation}

Evaluating \Eqn{R_Mu} at the inflection point using \Eqn{RI}, and rearranging yields 
\begin{equation}
q_I = \frac{\ews}{\ew}\hat\gkk_N \dot{\gkg}_I\;,
\end{equation}
thus
\begin{equation}
M = \frac{\ew m}{\ews \hat\gkk_N}\;.
\label{eq:MR}
\end{equation}
With these parameters determined, it is possible to calculate the pressure drop through a system when a non-Newtonian fluid is flowing within the system {\em a priori} as long as the flow is laminar and the inflection point assumption holds. 

In the tensor form, it is expected that the orientation of the resistance tensor will not change with flow conditions. Applying a Cholesky decomposition to the resistance tensor yields \cite{Dye_McClure_etal_13} 
\begin{equation}
\htR^w = \ten Q^w\vdot\ten \gkL^w\vdot\lrp{\ten Q^w}^{-1}\;,
\label{eq:Rdecomp}
\end{equation}
where the column vectors of $\ten Q^w$  are orthonormal eigenvectors, and $\ten\gkL^w$ is a diagonal tensor which contains the eigenvalues of $\hat{\ten{R}}^w$, which we will call $\gkl^w_i$, with $i=1,2,3$ corresponding to each of the primary directions of the coordinate system.  A similar Cholesky decomposition can be generated for $\Hat{\ten k}^w$ in \Eqn{ktR2}, denoted below
\begin{equation}
\htk^w = \ten Q^w_k\vdot\ten \gkL^w_k\vdot\lrp{\ten Q^w_k}^{-1}\;,
\end{equation}
where $\ten Q^w_k$ contains the orthonormal eigenvectors, with $\ten Q^w_k=\lrp{\ten Q^w_k}^{-1}$, and $\ten \gkL^w_k$ being a diagonal tensor containing the associated eigenvalues. Here, for the case of a Newtonian fluid flowing through the porous medium, the eigenvalues are $\hat{\gkk}_{Ni}$, and these are expected to be invariant with respect to fluid properties during laminar flow. The tensor form of the resistance model can thus be written as a direct extension of the scalar model as
\begin{equation} 
\gkL^w_i = \lrp{\gkL^w_\infty}_i + \frac{\lrp{\gkL^w_0}_i-\lrp{\gkL^w_\infty}_i}{1+\lrp{M_i v^\dol{w}}^N}\;,
\label{eq:aniR}
\end{equation}
where it is expected that $N = n$ as before, $M_i$ would be similarly calculated as in the isotropic case, and  
\begin{equation}
\lrp{\gkL^w_\infty}_i = \frac{\lrp{\ew}^2\hat{\mu}_\infty}{\hat\gkk_{Ni}} \;,
\lrp{\gkL^w_0}_i = \frac{\lrp{\ew}^2\hat{\mu}_0}{\hat\gkk_{Ni}}\;.
\label{eq:gkLi}
\end{equation}

It has been found that for the anisotropic case of transition flow in a porous medium, the inertial impacts on flow are only dependent on $v^\dol{w}$ \cite{Dye_McClure_etal_13}, thus it has been assumed here that the impact on non-Newtonian laminar flow is similarly only dependent on $v^\dol{w}$. The remaining independent quantities that are present in $\ten \gkL^w$ are assumed to be provided by the orientation of the tensor, which is assumed to be the same as for the Newtonian case; it is also assumed that $\ten Q^w$ in the non-Newtonian case is the same as the Newtonian case.

%%%%%%%%%%%%%%%%%%%%%%%%%%%%%%%%%%%%%%%%%%%%%%%

\section{Results and Discussion}

Because we have derived all equations as precise averages of micoscale quantities, integrated results from microscale analysis, computed using an analytical or numerical approach, can be used to evaluate and validate the results derived. This is necessary because some approximations have been made, as well as an assumed scaling ansatz. The validation of the results for non-Newtonian laminar flow will follow four different cases: (1) the case of three parallel slits, for which an analytical solution has been derived; (2) flow in a system of packed spheres, which will be simulated using OpenFOAM; (3) an anisotropic system of ellipsoids, which will also be simulated using OpenFOAM; and (4) flow through a rough slit geometry, for which simulation data has been published \cite{Zhang_Prodanovic_etal_19}. These four cases will be considered in turn in the subsections that follow.

\subsection{Parallel Slits}

\begin{figure}[b!]
    \centering
    \includegraphics[width=0.6\linewidth]{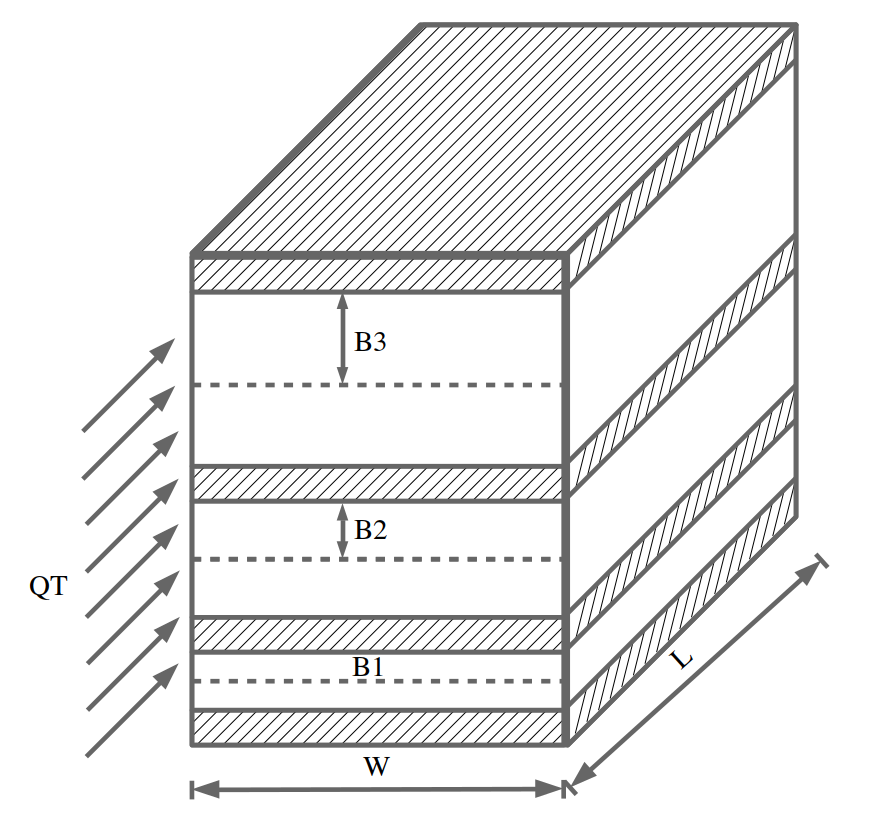}
    \caption{Illustration of a set of three slits aligned in the same principal direction of flow.}
    \label{fig:SLIT_DIA}
\end{figure}

It has become common practice to treat a porous medium either as a bundle of capillary tubes, or in some cases as a set of parallel slits. This is generally done because analytical solutions can often be derived for these simple geometries, and because similar physics are present i.e. a fluid interacting with a solid phase, flow through small pore throats, and averaging to a scale for which a general flow pattern may emerge. The system considered here consists of flow through a system of three parallel slits, as illustrated in Fig. \ref{fig:SLIT_DIA}. 

The relationship for flow of a Cross model fluid through a slit has been derived \cite{Sochi_15} and is
\begin{equation}
Q^\dol{w} = \frac{2WB^2I}{\lrp{\tau^{ws}}^2}\;,
\label{eq:Qslit}
\end{equation}
where $Q^\dol{w} = 2WB q^\dol{w}$ is the flow rate, $W$ is the width of the slit, $B$ is the half-height of the slit, $\tau^{ws}$ is the intrinsic average shear stress at the wall, and $I$ is an integral, given by
\begin{equation}
I = \ilims{0}{\dot{\gkg}^{ws}} \dot{\gkg}^2\lrp{\Hat{\mu}_\infty + \frac{\gkd}{1 + \lrp{m\dot{\gkg}}^n} }\lrp{\Hat{\mu}_\infty + \frac{\gkd}{1 + \lrp{m\dot{\gkg}}^n} -  \frac{n\gkd\lrp{m\dot{\gkg}}^n}{\lrb{1 + \lrp{m\dot{\gkg}}^n}^2} } \text{d}{\dot{\gkg}}\;,
\label{eq:Intslit}
\end{equation}
which evaluates to
\begin{equation}
I = \frac{\lrb{3\gkd^2(n-g_I)-\lrB{\gkd^2\lrp{n-3} + 2n\gkd\hat{\mu}_\infty }\lrp{g_I}^2 {}_2F_1\lrp{1,\frac 3n;1+ \frac 3n; -f_I} + 6n\gkd\hat{\mu}_\infty g_I + 2n\hat{\mu}_\infty^2 \lrp{g_I}^2 }\dot{\gkg}^3_w}{6 m \lrp{g_I}^2}\;,
\label{eq:Islit}
\end{equation}
where $\gkd$ is the difference between $\hat{\mu}_0$ and $\hat{\mu}_\infty$, $\dot{\gkg}^{ws}$ is the intrinsic average shear rate at the wall, ${}_2F_1$ is the real component of the hypergeometric function, and $f_I$ and $g_I$ are
\begin{equation}
f_I = m^n\dot{\gkg}^n_w \hbox{ and } g_I = 1 + f_I\;.
\end{equation}
The shear stress at the wall is given by
\begin{equation}
\tau^{ws} = \frac{B\gkD p^w}{L}\;,
\label{eq:tauw}
\end{equation}
where $L$ is the length of the slit and $\gkD p^w$ is the change in pressure across the system; and the shear rate at the wall is calculated numerically by finding the root of
\begin{equation}
\tau^{ws} = \hat{\mu}(\dot{\gkg}^{ws})\dot{\gkg}^{ws} = \lrb{\hat{\mu}_\infty + \frac{\hat{\mu}_0 - \hat{\mu}_\infty}{1 + \lrp{m\dot{\gkg}^{ws}}^n}}\dot{\gkg}^{ws}\;. 
\label{eq:tws}
\end{equation}

For slits in parallel, the flow rate is calculated for each individual slit and then the total flow rate is calculated by summing the flow rates. The system and fluid parameters used are listed in Table \ref{tab:Slits_Param}.
\begin{table}[t]
    \centering
    \caption{\label{tab:Slits_Param}System and fluid properties for test case of a set of parallel slits.}
    \begin{ruledtabular}
    \begin{tabular}{lccc}
         Parameter  & Slit 1 & Slit 2 & Slit 3  \\
         \cline{1-4}
         Half-height $B$ (m) \hspace{1cm} & $0.5\times 10^{-3}$ & $1.0\times 10^{-3}$ & $1.5\times 10^{-3}$ \\
         Length (m) & 0.01 & 0.01 & 0.01 \\
         Width (m) & 0.004 & 0.004 & 0.004 \\ 
         Porosity $\epsilon$ & 1.0 & 1.0 & 1.0 \\
         $\hat{\mu}_0$  (Pa$\cdot$s)      &  0.135 &  0.135 &  0.135  \\
         $\hat{\mu}_\infty$ (Pa$\cdot$s) \hspace{1cm}    &  0.00303 &  0.00303 &  0.00303 \\
         $m$ (s) & 0.0448 & 0.0448 & 0.0448 \\
         $n$ & 0.685 & 0.685 & 0.685 \\
         $\raw$ (kg/m$^3$) & 999.1 & 999.1 & 999.1 \\
    \end{tabular}
    \end{ruledtabular}
\end{table}
Shear stress and shear rate at the walls are calculated for each slit individually to calculate the flow rate. The hydraulic conductivity is then calculated using Darcy's law from the total flow rate $Q_T^\dol{w}$, and total inlet area $A_T^w$ using the below equation,
\begin{equation}
    \hat{K}^w_\text{Darcy} = \frac{Q_T^\dol{w}}{A_T^w}\lrp{\frac{1}{\raw g}\frac{\gkD \paw}{L}}^{-1} = \frac{1}{A_T^w}\lrp{Q_1^\dol{w} + Q_2^\dol{w} + Q_3^\dol{w}}\lrp{\frac{1}{\raw g}\frac{\gkD \paw}{L}}^{-1}\;,
\label{eq:KD}
\end{equation}
where $Q^\dol{w}_i$ is the flow rate for the $i$th slit calculated using the analytical solution corresponding to the geometric parameters of that slit. The change in pressure was the same for each slit. The computed hydraulic conductivity was then compared to the hydraulic conductivity calculated using the isotropic case of the derived theory, which can be written as %\Eqn{KtR2}
\begin{equation}
    \hat{K}^w_\text{TCAT} = -\frac{\lrp{\ew}^2 \raw g  v^{\dol{w}}} {\ews\hat{\mu}(\dot{\gkg}^{ws})\dot{\gkg}^{ws}}\;.
\label{eq:KsR1}
\end{equation}

\begin{figure}[t]
    \centering
    \begin{subfigure}[b]{0.49\linewidth}
        \includegraphics[width=\linewidth]{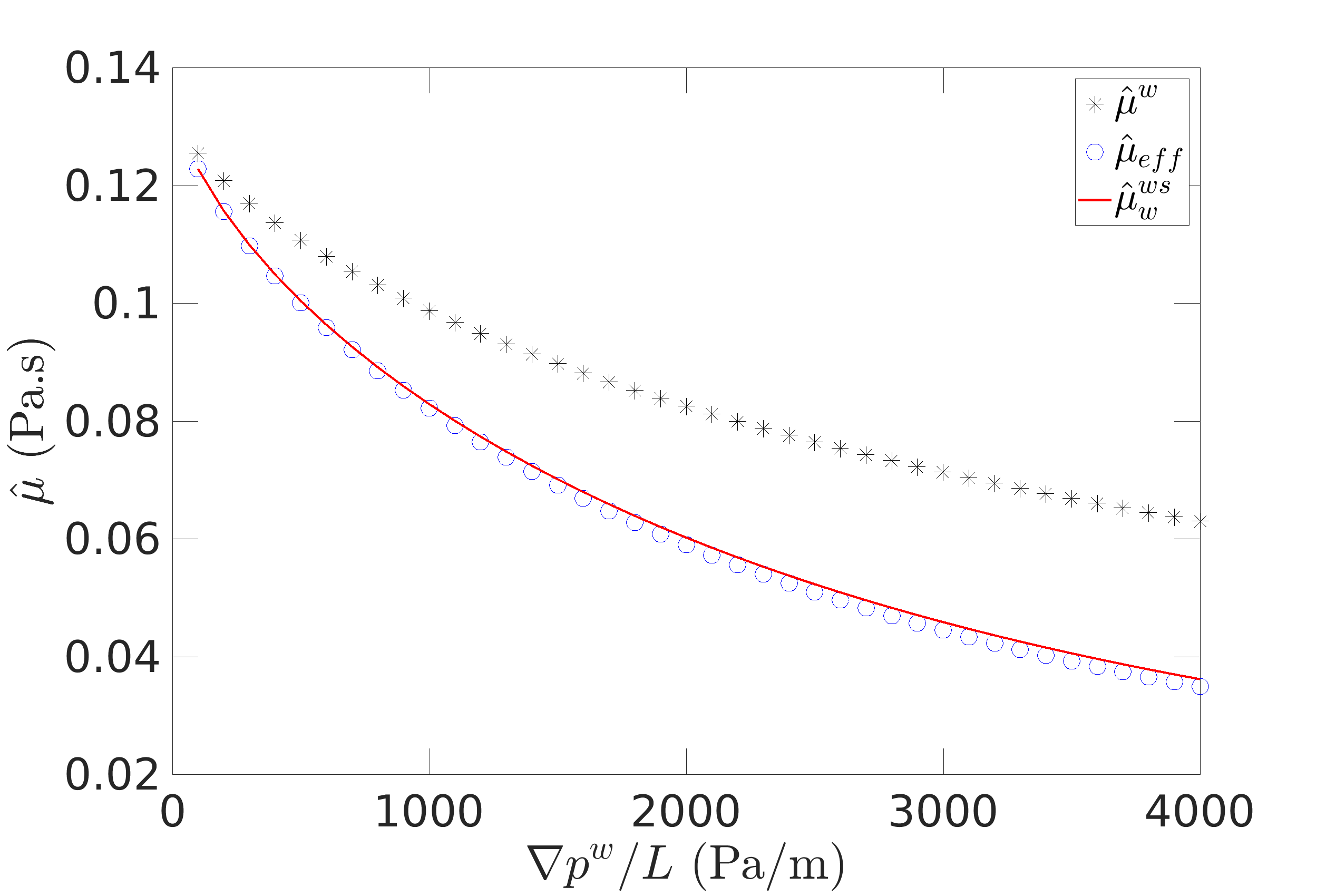}
        \caption{Series of Slits}
    \end{subfigure}
    \begin{subfigure}[b]{0.49\linewidth}
        \includegraphics[width=\linewidth]{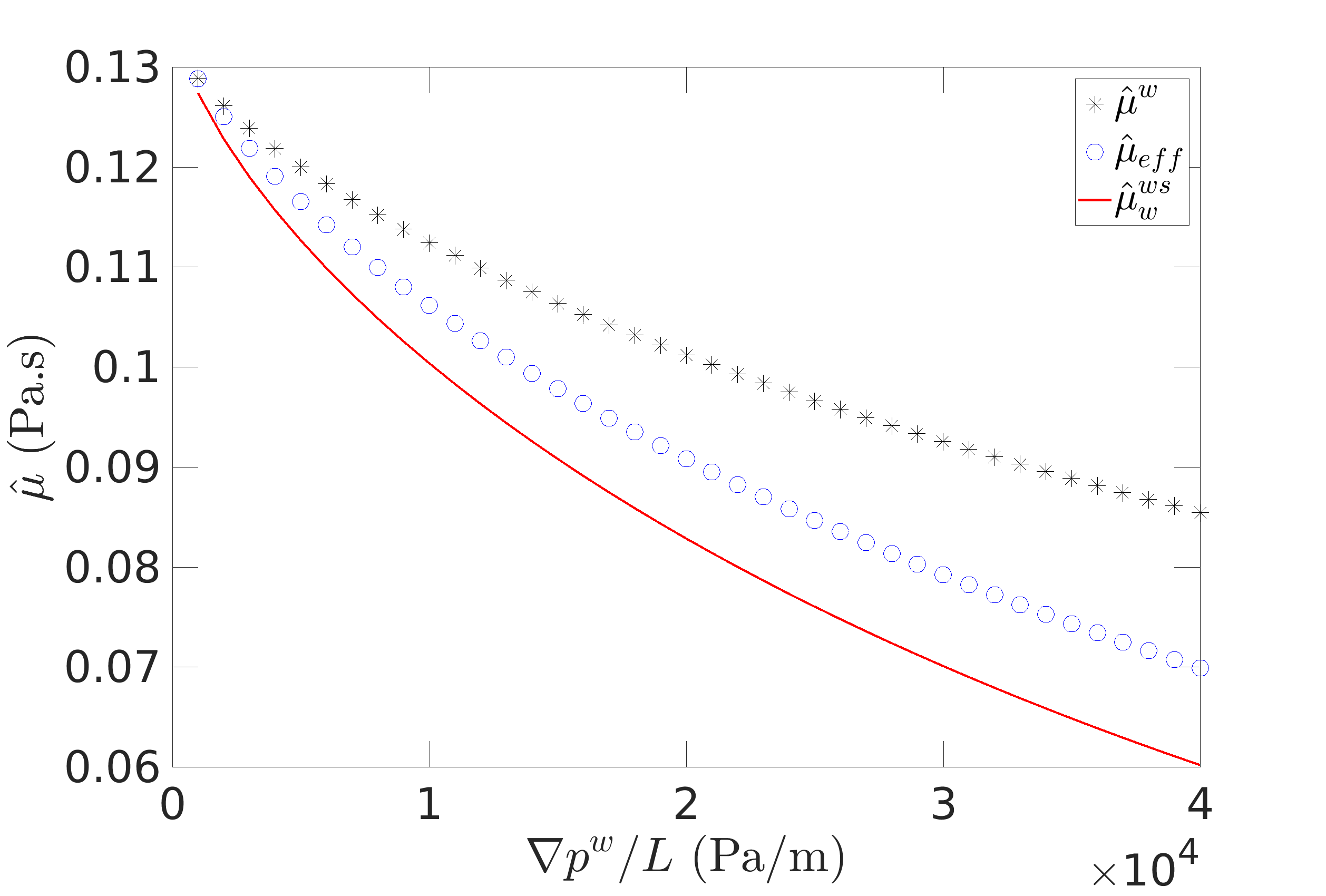}
        \caption{Single Slit}
    \end{subfigure} 
    \caption{(a) The bulk average, apparent, and interfacial viscosities measured in the series of slits of interest. (b) The bulk average, apparent, and interfacial viscosities measured in a single slit within the system of slits of interest, with a non-Newtonian fluid flowing through the system.}
    \label{fig:SLIT_VISC}
\end{figure}

\begin{figure}[t]
    \centering
    \begin{subfigure}[b]{0.49\linewidth}
        \includegraphics[width=\linewidth]{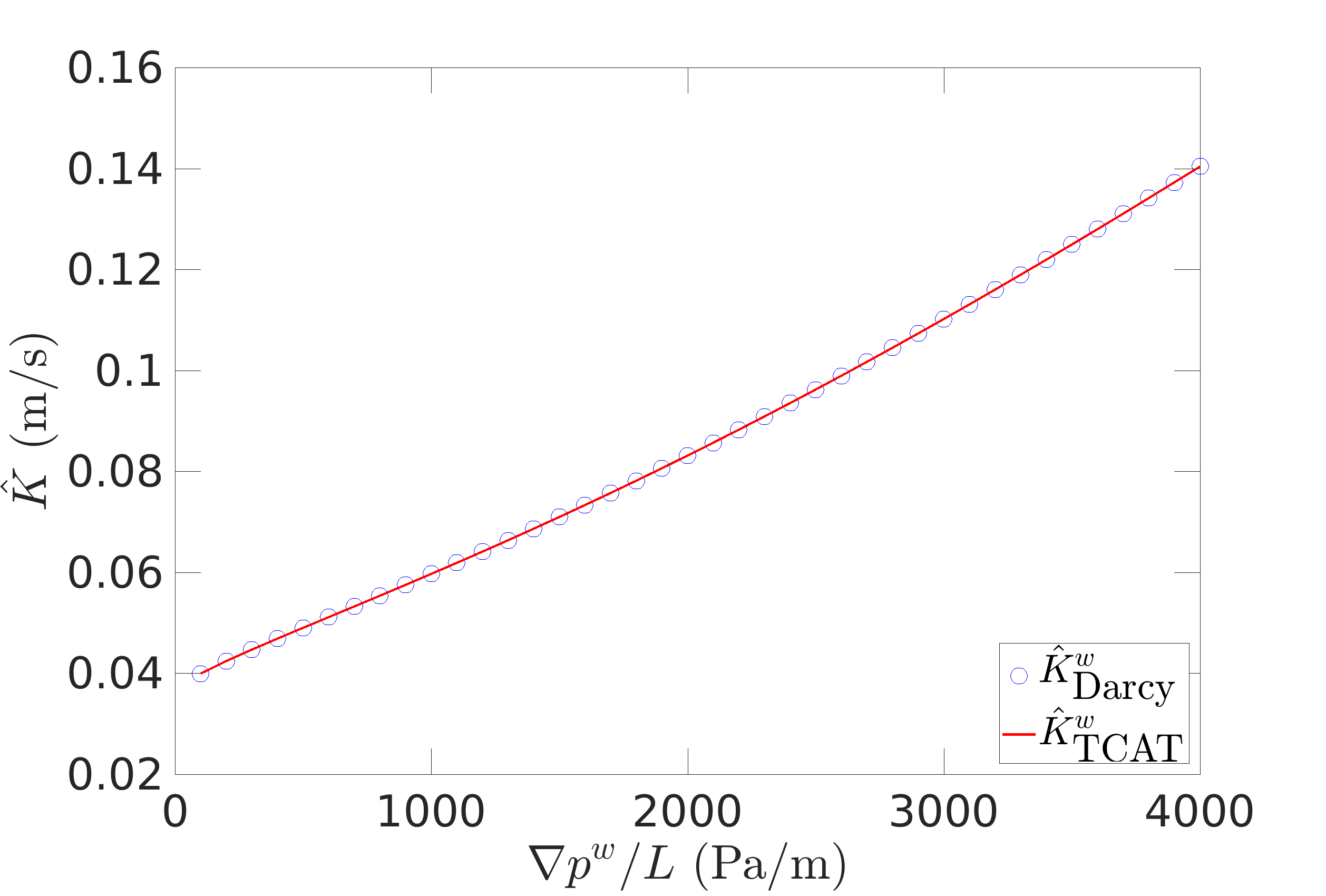}
        \caption{Hydraulic Conductivity}
    \end{subfigure}
    \begin{subfigure}[b]{0.49\linewidth}
        \includegraphics[width=\linewidth]{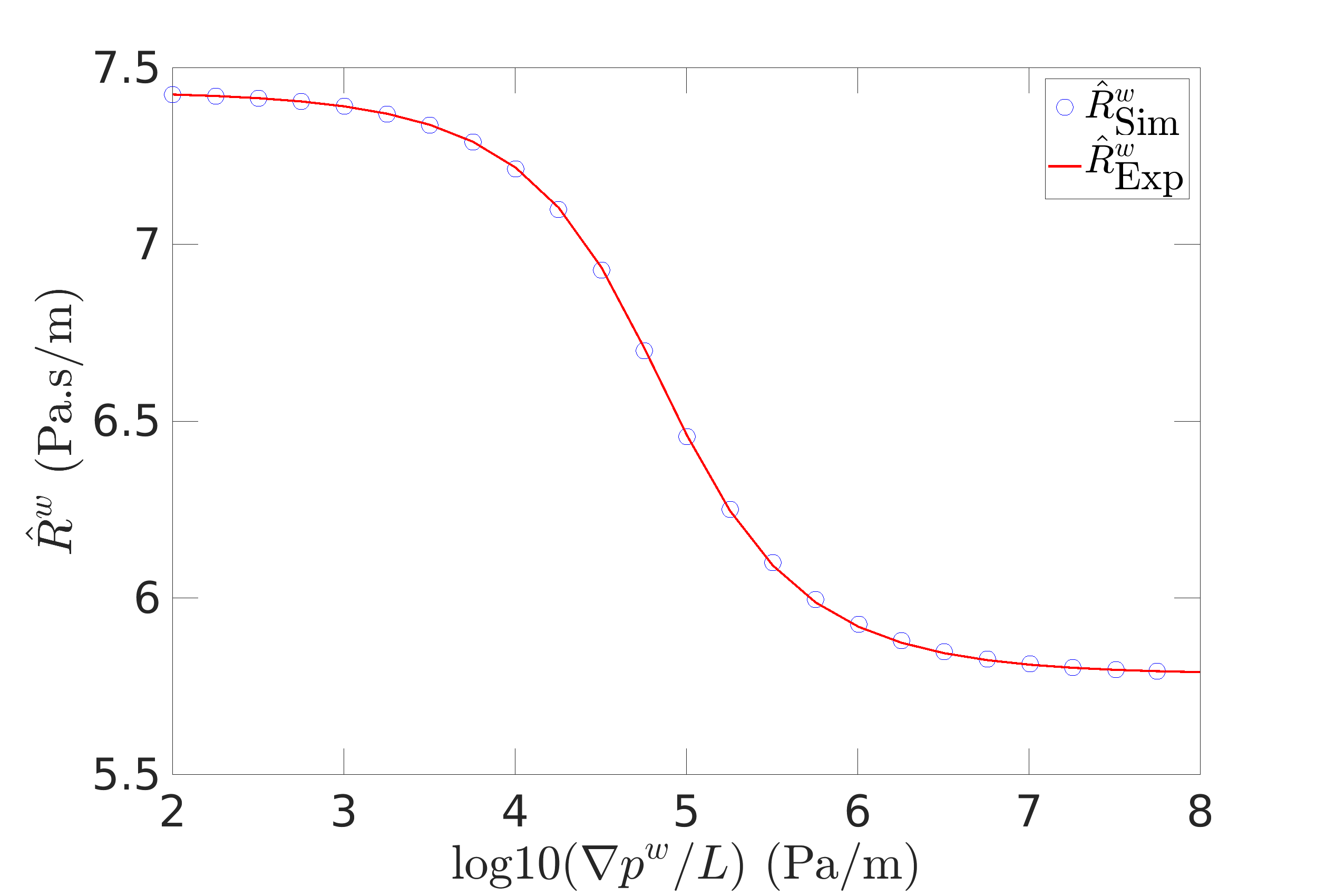}
        \caption{Hydraulic Resistance}
    \end{subfigure} 
    \caption{(a) Comparison of the Darcy calculated and TCAT calculated hydraulic conductivity. (b) Comparison of the hydraulic resistance calculated from simulation data and the expected resistance based on the proposed model.}
    \label{fig:COND_RES}
\end{figure}

In addition to the average viscosity at the fluid-solid interface being calculated, the bulk average viscosity $\hat{\mu}^w$, was calculated by integrating \Eqn{tws} over each slit, and the effective viscosity $\hat{\mu}_\text{eff}$ was calculated from \Eqn{muEff} to evaluate the traditional effective viscosity formulation. These three different viscosities are shown in Fig. \ref{fig:SLIT_VISC}(a) for the full set of slits, and Fig. \ref{fig:SLIT_VISC}(b) for a single slit within the set. While it may appear that the effective viscosity describes the average viscosity at the $ws$-interface based on Fig. \ref{fig:SLIT_VISC}(a), it can be observed from the single-slit results shown in Fig. \ref{fig:SLIT_VISC}(b) that this is a fortuitous result. Additionally, it may be seen from Fig. \ref{fig:COND_RES}(a) that the hydraulic conductivity calculated using \Eqn{KsR1} yields an exact result for laminar flow through a domain consisting of a set of slits based upon comparison with the analytical solution.
 
The resistance function, which was proposed in \Eqn{RES_FORM}, and the {\em a priori} calculations of its parameters were tested for this case of three slits. The parameters were calculated according to \Eqn{RI0} through \Eqn{MR}, and these were compared to the output hydraulic resistance, calculated from \Eqn{KsR1} and the conversion present in \Eqn{KtR2}. The results of this can be seen in Fig. \ref{fig:COND_RES}(b). Based on these results, the  parameter estimation procedure described above is validated for this case, allowing for prediction of the pressure drop during non-Newtonian flow in a system without requiring simulation or experimental data for non-Newtonian flow. 

\subsection{Packed Spheres}

\begin{figure}[t!]
    \centering
    \begin{subfigure}[b]{0.49\linewidth}
        \includegraphics[width=\linewidth]{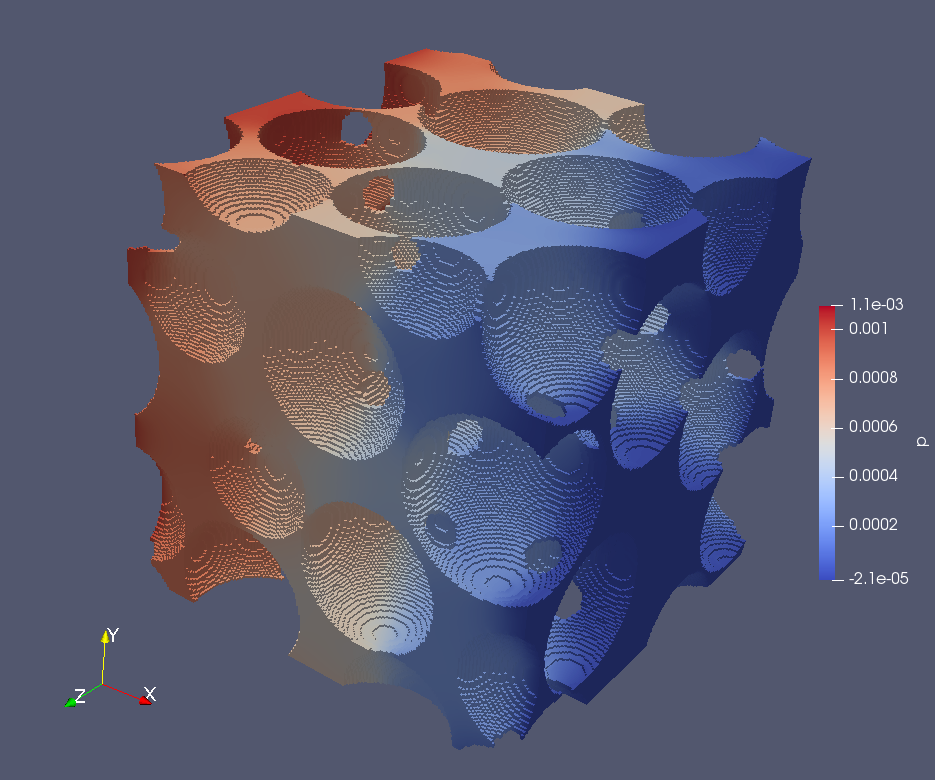}
        \caption{Fluid domain.}
    \end{subfigure}
    \begin{subfigure}[b]{0.49\linewidth}
        \includegraphics[width=\linewidth]{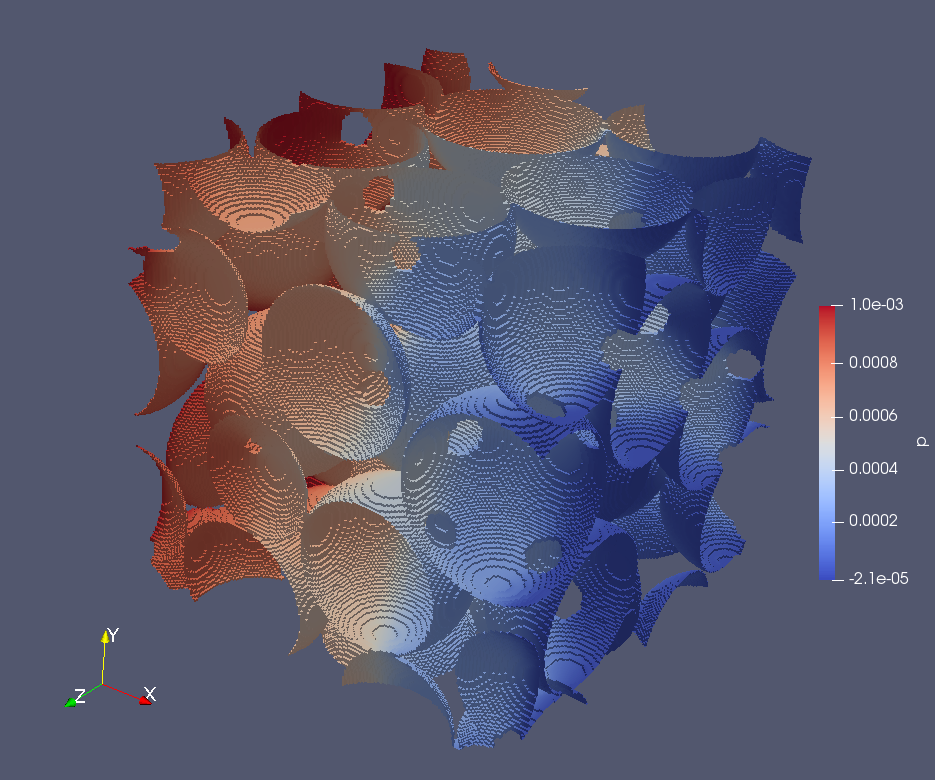}
        \caption{Solid-surface domain.}
    \end{subfigure} 
    \caption{(a) Fluid domain of the system of packed spheres. (b) Solid surface domain of the system of packed spheres. In both cases coloration is generated from pressure data.}
    \label{fig:OpenFOAM_Domains}
\end{figure}

The second case considered for evaluating and validating the non-Newtonian TCAT model is a set of randomly packed spheres. This system is not large enough to constitute a representative elementary volume for macroscale parameters pertaining to packed spheres, but the theory developed here is still valid. The system and fluid parameters used are listed in Table \ref{tab:OF_Param}. The domain was square and discretized with 50 cells in each dimension. The domain generated is shown in Fig. \ref{fig:OpenFOAM_Domains} (a). It was also necessary to calculate shear rates and dynamic viscosities at the fluid-solid interface; Paraview was used to isolate cells containing an interface \cite{Ayachit_19}, which is shown in Fig. \ref{fig:OpenFOAM_Domains} (b). The OpenFOAM post-processing function wallShearStress was also used \cite{Greenshields_18}. In Paraview, quantities were integrated and normal vectors for the cells at the fluid-solid interface were calculated \cite{Ayachit_19}.

\begin{table}[t]
    \centering
    \caption{\label{tab:OF_Param} System and fluid properties for packed spheres test case.}
    \begin{ruledtabular}
    \begin{tabular}{lc}
         Number of Spheres & 25 \\
         Domain Length (m) & 0.001 \\
         Porosity (m$^3$/m$^3$) & 0.399165 \\
         Mean Radius (m) & $1.787\times 10^{-4}$ \\
         Log Normal Radius Variance \hspace{1cm} & 0.004 \\
         Number of Cells & $50^3$ \\
         $\hat{\mu}_0$  (Pa$\cdot$s)      &  0.135  \\
         $\hat{\mu}_\infty$ (Pa$\cdot$s) \hspace{1cm}    &  0.00303 \\
         $m$ (s) & 0.0448 \\
         $n$ & 0.685 \\
         $\raw$ (kg/m$^3$) & 999.1 \\
    \end{tabular}
    \end{ruledtabular}
\end{table}

The case of packed spheres is one in which the assumption that the pressure deviation term is negligible, which was made between \Eqn{Ta5} and \Eqn{Ta6}, is no longer valid. This is because the pressure fluctuation term is correlated with the direction of the outward normal vector for each sphere. This correlation is expected occur in generally for unconsolidated media, and may be expected to arise for other simulated systems as well; for systems such as slits this force can be neglected. The scalar form of \Eqn{TR1} would be
\begin{equation}
    \hsR^w v^\dol{w} = |\ieT ws |\;,
\label{eq:RD_RT1}
\end{equation}
where $|\ieT ws|$ will have a magnitude directed in a single principal direction. The scalar form of the momentum transfer term, based on \Eqn{Ta5} and assuming that the viscosity fluctuation term is negligible but the pressure fluctuation term is not, is
\begin{equation}
    |\ieT ws| = |\big\langle\lrp{\piw-p_w^\ws}\vnmi w \big\rangle_{\Dm\ws,\Dm\ws}| - \ews\hat{\mu}_w^{ws}\dot{\gkg}^{\dol{ws}}_w\;,
\label{eq:RD_Ta1}
\end{equation}
where $|\big\langle\lrp{\piw-p_w^\ws}\vnmi w \big\rangle_{\Dm\ws,\Dm\ws}|$ will have a magnitude directed in a single principal direction. Assuming that the principal direction of the packed sphere system is the x-direction, while vector quantities directed in the other directions are negligible, and substituting \Eqn{RD_Ta1} into \Eqn{RD_RT1}, the hydraulic resistance for this system may be calculated by
\begin{equation}
\hsR^w = \frac{1}{v_x} \lrp{|\big\langle\lrp{\piw-p_w^\ws}\vnmi w \big\rangle_{\Dm\ws,\Dm\ws}|_x - \ews\hat{\mu}_w^{ws}\dot{\gkg}^{\dol{ws}}_w}\;.
\label{eq:RD_Rw1}
\end{equation}
The hydraulic conductivity may be calculated both from Darcy's law, and from the TCAT formulation. The Darcy's law form of the hydraulic conductivity is calculated by 
\begin{equation}
    \hat{K}^w_\text{Darcy} = q^\dol{w}\lrp{\frac{1}{\raw g}\frac{|\del \paw|}{L}}^{-1}\;,
\label{eq:RD_KwD2}
\end{equation}
while the TCAT formulation of the hydraulic conductivity was calculated from the scalar form of \Eqn{KtR2} 
\begin{equation}
{\hsK^w}_\text{TCAT} = \frac{\lrp{\ew}^2\rho g}{\hsR^w}\;.
\label{eq:RD_Kw2}
\end{equation}

The porosity used was calculated from $\ew = q^\dol{w}/v^\dol{w}$, where $q^\dol{w}$ was calculated by dividing the integrated flow rate at the inlet/outlet of the system, calculated using utilities in OpenFOAM, by the total domain area at the inlet and outlet, and $v^\dol{w}$ was calculated by direct averaging of microscale data using Paraview. 

\begin{figure}[t!]
    \centering
    \begin{subfigure}[b]{0.49\linewidth}
        \includegraphics[width=\linewidth]{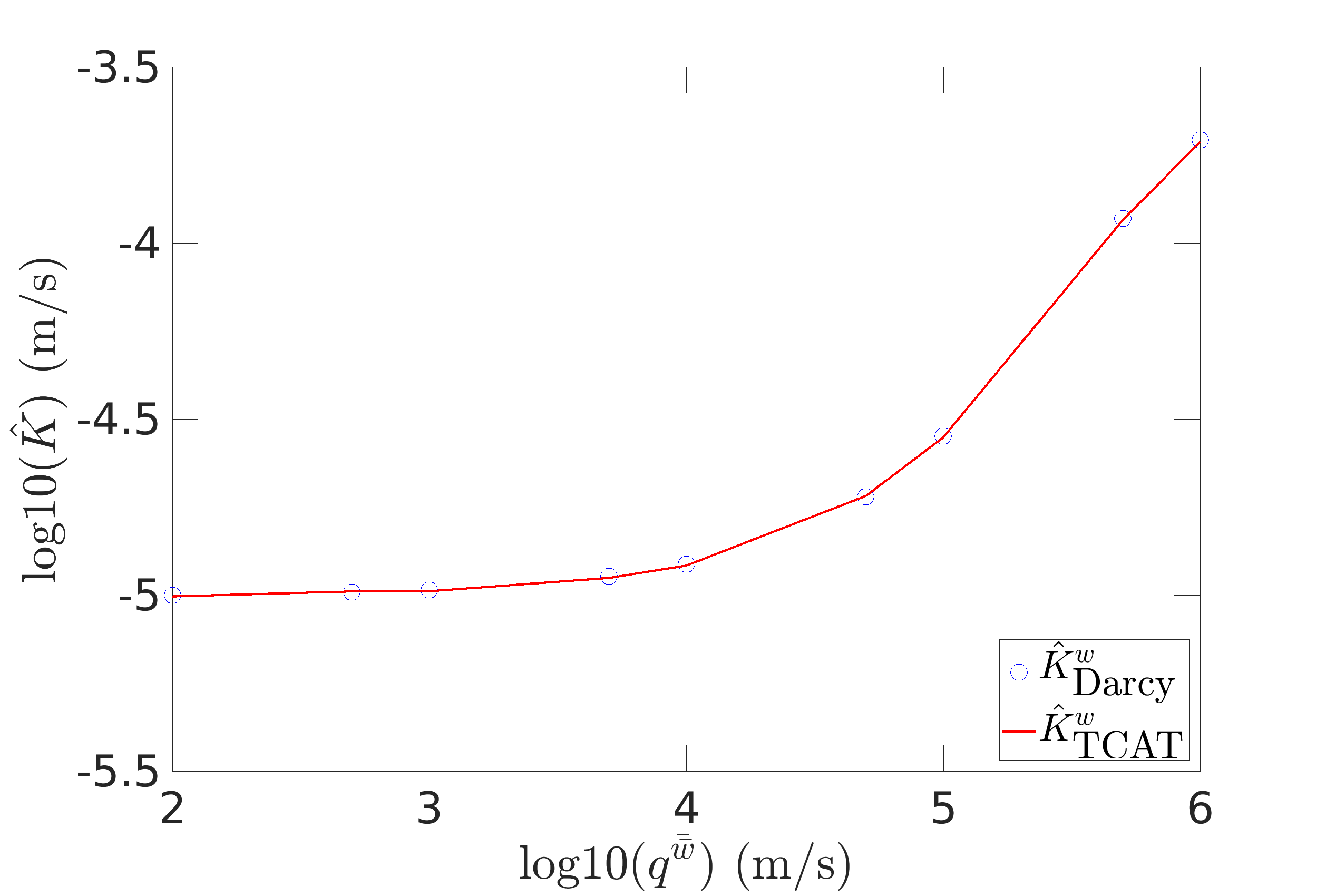}
        \caption{Hydraulic Conductivity}
    \end{subfigure}
    \begin{subfigure}[b]{0.49\linewidth}
        \includegraphics[width=\linewidth]{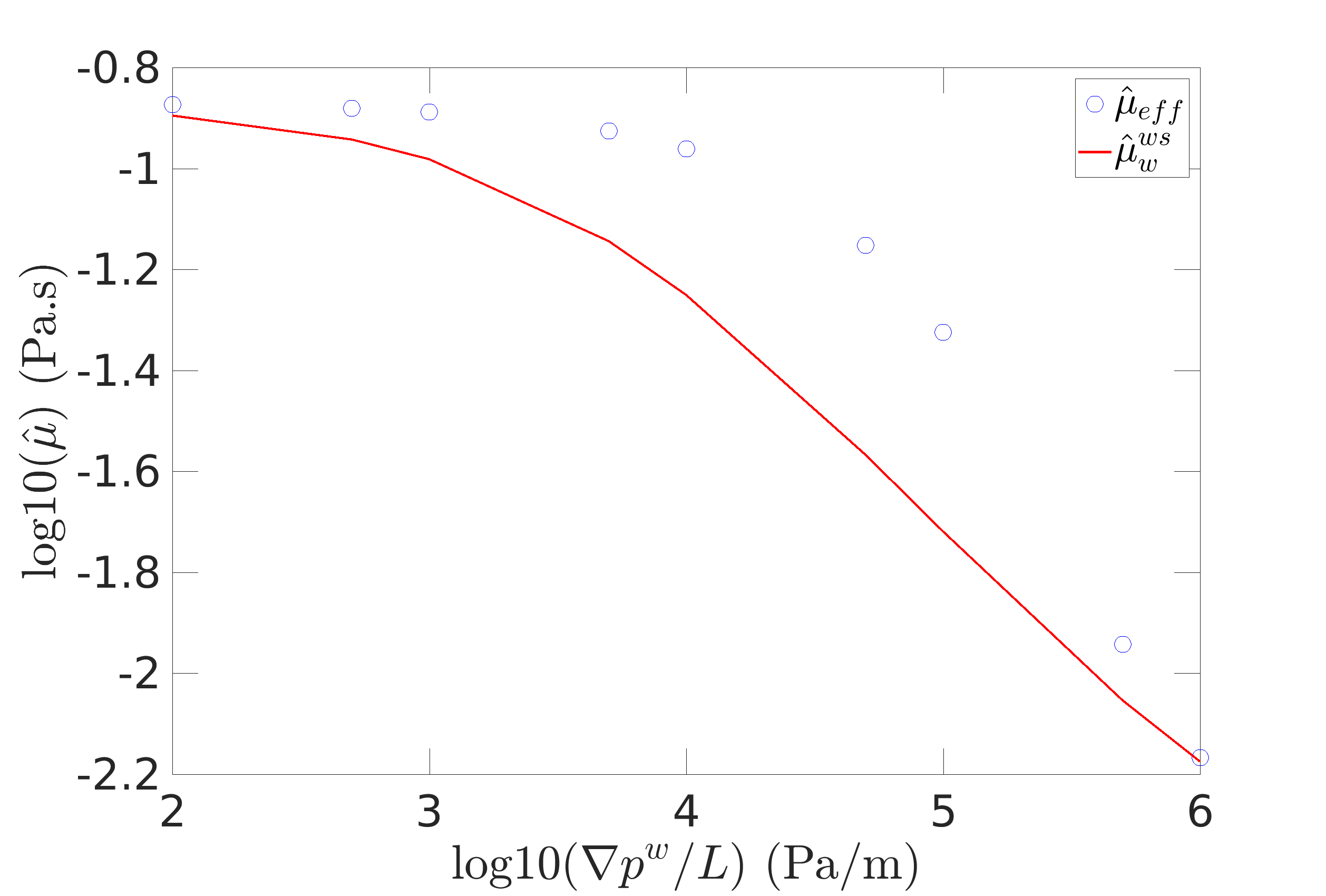}
        \caption{Characteristic Viscosities}
    \end{subfigure} \\
    \begin{subfigure}[b]{0.49\linewidth}
        \includegraphics[width=\linewidth]{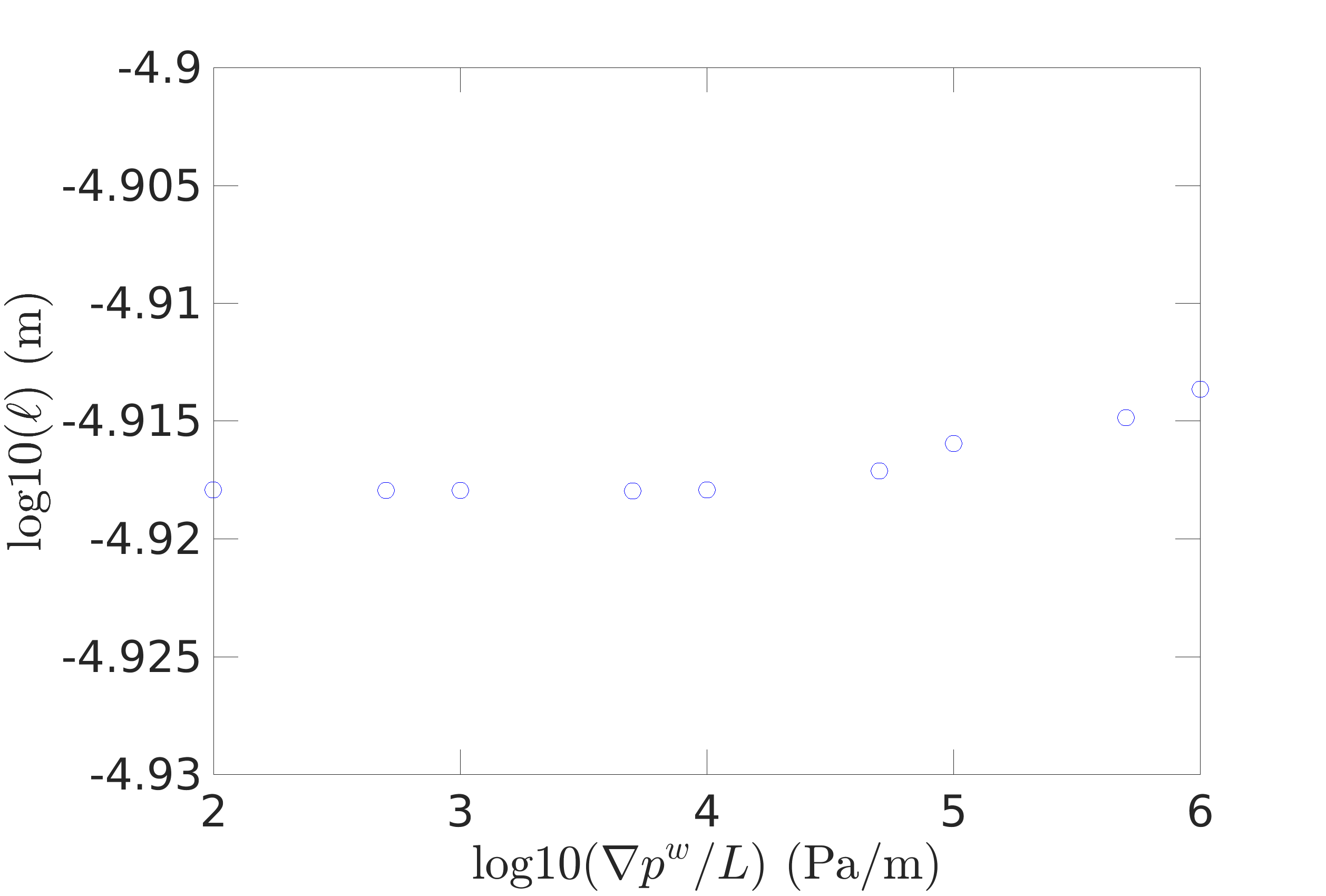}
        \caption{Characteristic Length}
    \end{subfigure} 
    \caption{Comparison of model coefficients for the spherical test case: (a) Darcy and TCAT calculated hydraulic conductivity; (b) effective viscosity and $\hat{\mu}^{ws}_w$; and (c) characteristic length.}
    \label{fig:SPHERE_COND_VISC_LENGTH}
\end{figure}

A comparison between the hydraulic conductivity calculated using \Eqn{RD_KwD2} and \Eqn{RD_Kw2} is shown in Fig. \ref{fig:SPHERE_COND_VISC_LENGTH}(a)---showing excellent agreement.  A relative error between these two calculations was computed using
\begin{equation}
\text{error}(\%) = \frac{|x_\text{approx} - x_\text{actual} |}{x_\text{actual}} \times 100\;,
\label{eq:RelError}
\end{equation}
where $x_\text{actual}$ is the known solution, in this case $\hat{K}^w_\text{Darcy}$. The relative error for this calculation was low, averaging 0.64\%, and with a maximum error of 1.47\% at the highest flow rate.

In addition to comparing the hydraulic conductivity calculated using Darcy's law and the theory proposed here, the effective viscosity calculated using \Eqn{muEff} and the microscale averaged viscosity $\hat{\mu}^{ws}_w$ were compared, which is shown in Fig. \ref{fig:SPHERE_COND_VISC_LENGTH}(b). It is apparent that the effective viscosity is not comparable to the viscosity at the $ws$-interface, providing another example of the deficiency of the standard analysis approach. To predict the resistance for this system, an approximation for the pressure deviation forces is required. This was proposed earlier in \Eqn{pf} and was calculated using 
\begin{equation}
    \ell = \frac{\big\langle\lrp{\piw-p_w^\ws}\vnmi w \big\rangle_{\Dm\ws,\Dm\ws}}{|\del\paw|}\;,
\end{equation}
where $\ell$ is the characteristic length of interest. This was a near constant value for the packed sphere system, as can be seen from Fig. \ref{fig:SPHERE_COND_VISC_LENGTH}(c).

The characteristic length is the key to predicting the hydraulic resistance {\em a priori}, as this allows the prediction of both the deviation pressure term and the viscosity at the fluid-solid interface. This characteristic length is of the same order of magnitude as the Sauter mean diameter, and it is expected that it will be dependent on the geometry of the system.

\subsection{Ellipsoids}

\begin{figure*}[t!]
    \centering
    \begin{subfigure}[b]{0.49\linewidth}
        \includegraphics[width=\linewidth]{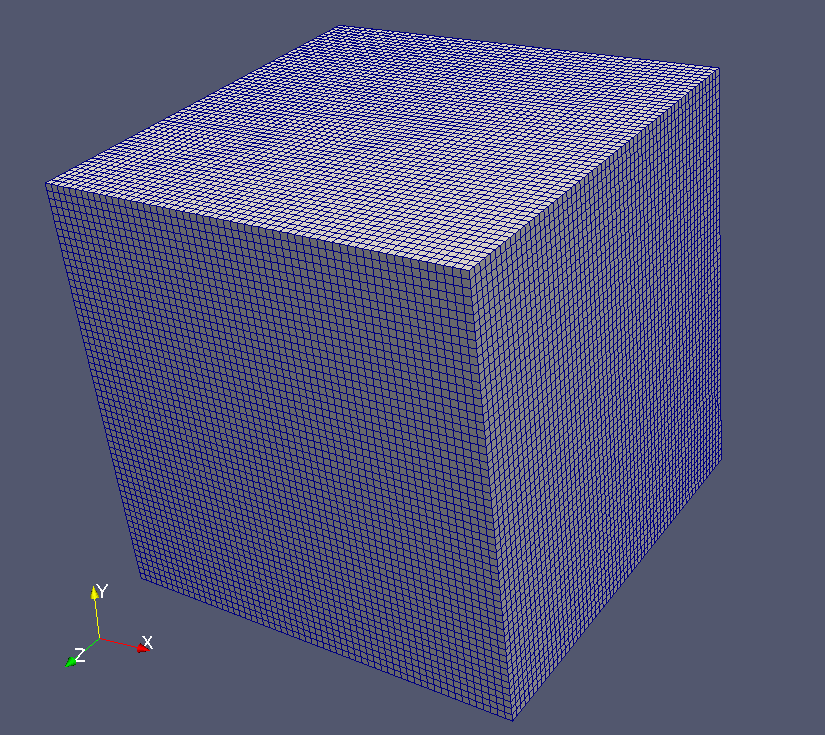}
        \caption{Fluid domain.}
    \end{subfigure}
    \begin{subfigure}[b]{0.49\linewidth}
        \includegraphics[width=\linewidth]{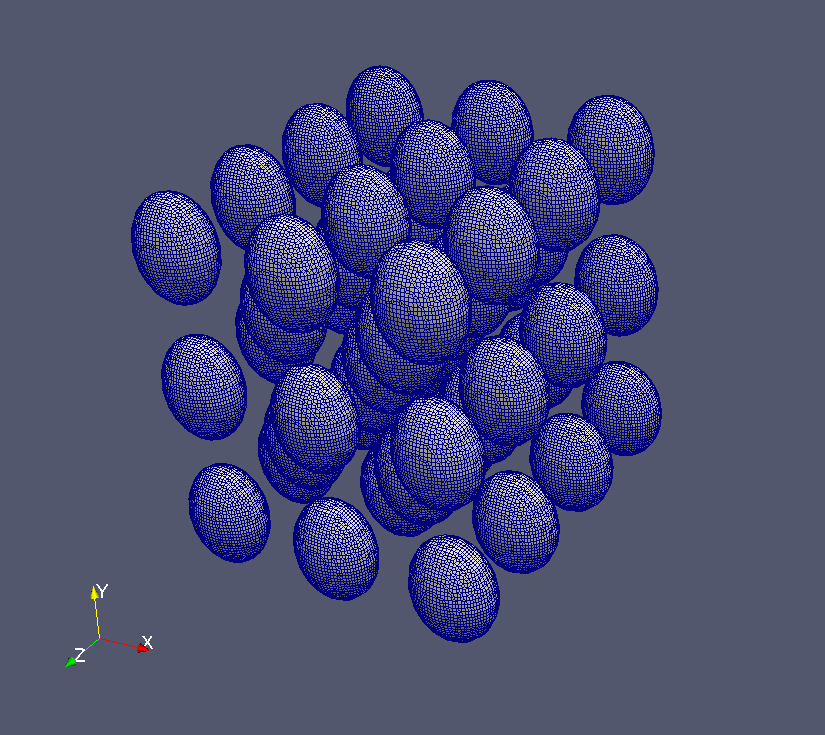}
        \caption{Solid-surface domain.}
    \end{subfigure} 
    \caption{(a) Fluid domain of the system of ellipsoids. (b) Solid surface domain of the system of ellipsoids.}
    \label{fig:OF_Ellips}
\end{figure*}

The third evaluation and validation case considered is a set of ellipsoids for which the principal directions are aligned with the direction of flow to simplify the analysis without loss of generality. This case allows for an investigation of non-Newtonian flow in an anisotropic porous medium system, which has received scant attention in the literature using any approach \cite{Fadili_Tardy_etal_02,Orgeas_Idris_etal_06,Di-Federico_Pinelli_etal_10}. The ellipsoid system was generated using the software FreeCAD, an STL mesh file was generated, and the snappyHexMesh utility was used to generate the mesh and surface shown in Fig. \ref{fig:OF_Ellips}(a) and (b).

Before simulated flow of a non-Newtonian fluid through the system was undertaken, simulation of a Newtonian fluid was conducted to determine the baseline eigenvectors and eigenvalues for the system. A pressure drop was applied across opposing faces in one-dimension at a time, ranging from 100 to 500 Pa/m, applying peridic boundary conditions to the faces corresponding to the other two dimensions in each case. The ellipsoids were arranged such that they did not intersect the faces of the system, allowing for comparison between pressure drops in each dimension without apparent entry or exit effects. The average velocity was calculated after completion of each simulation using the variable integration utility in Paraview. Using the average velocity and pressure drop data, the resistance tensor was fit to the equation
\begin{equation}
\htR^w\vdot\vaw = \del\paw\;,
\end{equation}
using the linsolve utility in Matlab. The eigenvalue decomposition utility in Matlab was then used to calculate the eigenvectors and eigenvalues of the resistance tensor such that $\hat{\ten R}^w = \ten Q^w\vdot\ten \gkL^w\vdot\lrp{\ten Q^w}^{-1}$. The eigenvector matrix was very near the identity tensor, as would be expected for geometric alignment with the principal directions of flow, but was slightly off due to the approximation required for the snapping procedure of the snappyHexMesh utility. To determine the goodness of fits, the change in pressure calculated using the fitted resistance tensor was compared to the actual change in pressure computed from the simulations, and the relative error was calculated using \Eqn{RelError}. This produced an average relative error less than 1\%, with a maximum relative error of 1.5\%. 

Simulations of a non-Newtonian fluid flowing through this same ellipsoid system were carried out next, using the same fluid parameters that were used in the non-Newtonian packed spheres case, described in Table \ref{tab:OF_Param}. The simulations were carried out similarly to the Newtonian flow simulations, with pressure drops ranging from 100 to 50,000 Pa/m in each dimension, and periodic boundaries in each of the other two dimensions. Average velocities were calculated using Paraview similarly to the Newtonian case. A set of resistance eigenvalues were calculated from 
\begin{equation} 
\gkL^w_i = \frac{\lrb{\lrp{\ten Q^w}^{-1}\cdot\vaw }_i}{\lrb{\ten Q^w\cdot\lrp{\del\paw}}_i },
\label{eq:LambEff}
\end{equation}
where $\gkL^w_i$ is the $ii$th entry of the diagonal tensor $\ten \gkL^w$, and $\ten Q^w$ is composed of the same set of eigenvectors as the Newtonian case. A non-Newtonian resistance model was fit to this data using \Eqn{aniR}. It was found that each of the resistance eigenvalues were dependent on the magnitude of the velocity, rather than the velocity in each dimension, similarly to cases where the inertial resistance was investigated \cite{Dye_McClure_etal_13}. Additionally, it was found that the resistance in each dimension could be calculated using the magnitude of the velocity and a scaling associated with each dimension, i.e., there existed an invariant tensor $\ten{\sol{\gkL}}^w$ such that 
\begin{equation}
\ten\gkL^w = \gkl^w_c\lrp{v^\dol{w}} \ten{\sol{\gkL}}^w,
\end{equation}
where $\gkl^w_c\lrp{v^\dol{w}}$ is some characteristic eigenvalue quantity that is a function of $v^\dol{w}$. We chose a characteristic eigenvalue that is the trace of the effective eigenvalue tensor, $\gkl^w_c = \text{tr}\lrp{\ten \gkL^w}$. Here we will call the invariant tensor, $\ten{\sol{\gkL}}^w = \frac{1}{\gkl^w_c}\ten\gkL^w$, the resistance orientation tensor, and this was found to be the same as the resistance orientation tensor of the Newtonian fluid when calculated in the same way. This greatly reduces the complexity of the calculations for non-Newtonian fluid flow in an anisotropic porous media, as the eigenvectors and the eigenvalue orientation tensor only need to be calculated in the Newtonian case, with the non-Newtonian case only requiring calculation of this characteristic eigenvalue to be predictive. Calculating the trace of the eigenvalue tensor for each simulation, the functional form of $\gkl^w_c$ was found to be 
\begin{equation} \label{eq:Lc}
\gkl^w_c\lrp{v^\dol{w}} = \gkl_\infty = \frac{\gkl_0 - \gkl_\infty}{1+\lrp{M v^\dol{w}}^N}\;,
\end{equation}
where $N$ is the same as the fluid parameter $n$, $M$ is fitted to this data, and $\gkl_\infty$ and $\gkl_0$ are the infinite and zero velocity resistance eigenvalues, which can be calculated from the Newtonian flow data and the non-Newtonian fluid parameters. To calculate $\gkl_\infty$ and $\gkl_0$, the Newtonian permeability eigenvalues must first be calculated from 
\begin{equation}
    \lrp{\ew}^2\hat{\mu}_N\lrp{\ten\gkL^w_N}^{-1} = \lrp{\ew}^2
    \begin{bmatrix} 
        \frac{\hat{\mu}_N}{\lrp{\gkl_N}_1} & 0 & 0 \\
        0 & \frac{\hat{\mu}_N}{\lrp{\gkl_N}_2} & 0 \\
        0 & 0 & \frac{\hat{\mu}_N}{\lrp{\gkl_N}_3}
    \end{bmatrix} = \lrp{\ew}^2
    \begin{bmatrix}
        \lrp{\hat{\gkk}_N}_1 & 0 & 0 \\
        0 & \lrp{\hat{\gkk}_N}_2 & 0 \\
        0 & 0 & \lrp{\hat{\gkk}_N}_3
    \end{bmatrix}\;, 
\end{equation}
where $\hat{\mu}_N$ is the viscosity of the Newtonian fluid. $\gkl_\infty$ and $\gkl_0$ are then calculated from the equations below,
\begin{equation}
    \gkl_\infty = \lrp{\ew}^2\lrb{ \frac{\hat{\mu}_\infty}{\lrp{\hat{\gkk}_N}_1} + \frac{\hat{\mu}_\infty}{\lrp{\hat{\gkk}_N}_2} + \frac{\hat{\mu}_\infty}{\lrp{\hat{\gkk}_N}_3} }
    \hspace{0.75cm},\hspace{0.75cm}
    \gkl_0 = \lrp{\ew}^2\lrb{\frac{\hat{\mu}_0}{\lrp{\hat{\gkk}_N}_1} + \frac{\hat{\mu}_0}{\lrp{\hat{\gkk}_N}_2} + \frac{\hat{\mu}_0}{\lrp{\hat{\gkk}_N}_3}}\;.
\end{equation}

\begin{figure}[t!]
    \centering
    \includegraphics[width=0.7\linewidth]{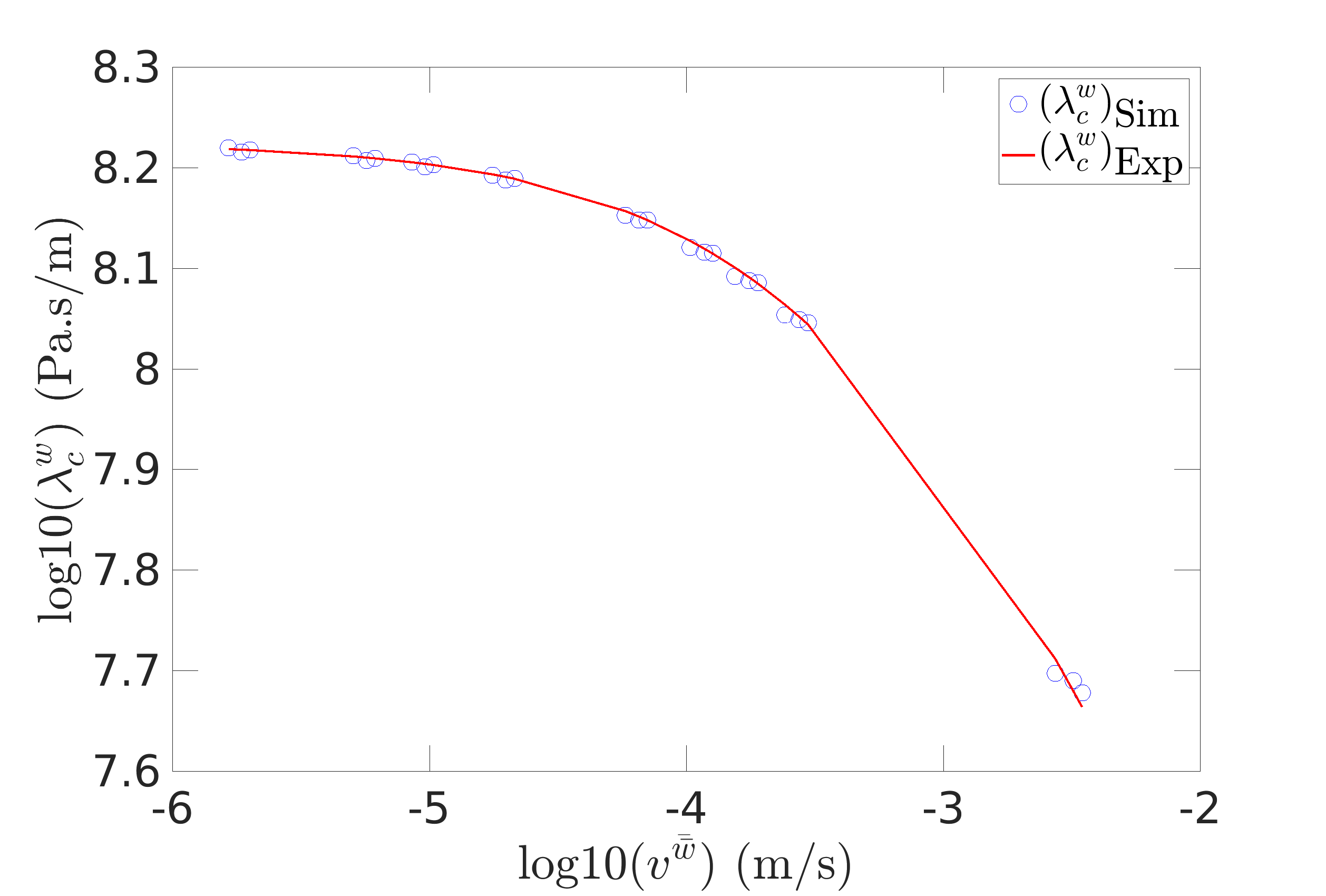}
    \caption{The characteristic eigenvalues for the resistance experienced during non-Newtonian flow through a set of ellipsoids and their expected values based on the proposed model.}
    \label{fig:ELLIPSE_EIG}
\end{figure}

From the simulation data the effective characteristic eigenvalue, calculated from the trace of that eigenvalue vector as described above, is compared to the characteristic resistance eigenvalue calculated from \Eqn{Lc} in Fig. \ref{fig:ELLIPSE_EIG}. This validation shows that this characteristic eigenvalue equation may be used with the eigenvalue orientation tensor and eigenvectors from the Newtonian case to predict flow of the non-Newtonian fluids through the system. 

As with the packed sphere case, fluctuation compression forces were present for the ellipsoid system. These were calculated by averaging at the microscale using Paraview, as in the packed sphere case. It was found here that a characteristic length tensor $\ten L$ existed such that 
\begin{equation}
    \big\langle\lrp{\piw-p_w^\ws}\vnmi w \big\rangle_{\Dm\ws,\Dm\ws} \approx \ten L\cdot\del\paw\;.
\end{equation}

It was possible to calculate this characteristic length tensor by using the geometric orientation tensor $\ten G^s$, calculated from
\begin{equation}
    \ten G^s = \big\langle\vnmi s\vnmi s \big\rangle_{\Dm\ws,\Dm\ws}\;,
\end{equation}
where $\vnmi s$ is the outward normal to the solid phase of the system, which was calculated at the microscale in Paraview. It is also possible to carry out an eigenvalue decomposition on $\ten G^s$ such that
\begin{equation}
\ten G^s = \ten Q^s\cdot\ten\gkL^s\cdot\lrp{\ten Q^s}^{-1},
\end{equation}
where $\ten Q^s$ is a tensor composed of the eigenvectors of $\ten G^s$ and $\ten\gkL^s$ is a diagonal tensor whose entries include the eigenvalues of $\ten G^s$. It was found that 
\begin{equation}
    \ten L \approx \frac{d_S}{2\pi} \ten Q^s\cdot\sqrt{\ten \gkL^s}\cdot\lrp{\ten Q^s}^{-1},
\end{equation}
where $d_S$ is the Sauter mean diameter, which is equal to 
\begin{equation}
    d_S = 6\frac{\ew}{\ews}.
\end{equation}

\begin{figure}[t!]
    \centering
    \begin{subfigure}[b]{0.329\linewidth}
        \includegraphics[width=\linewidth]{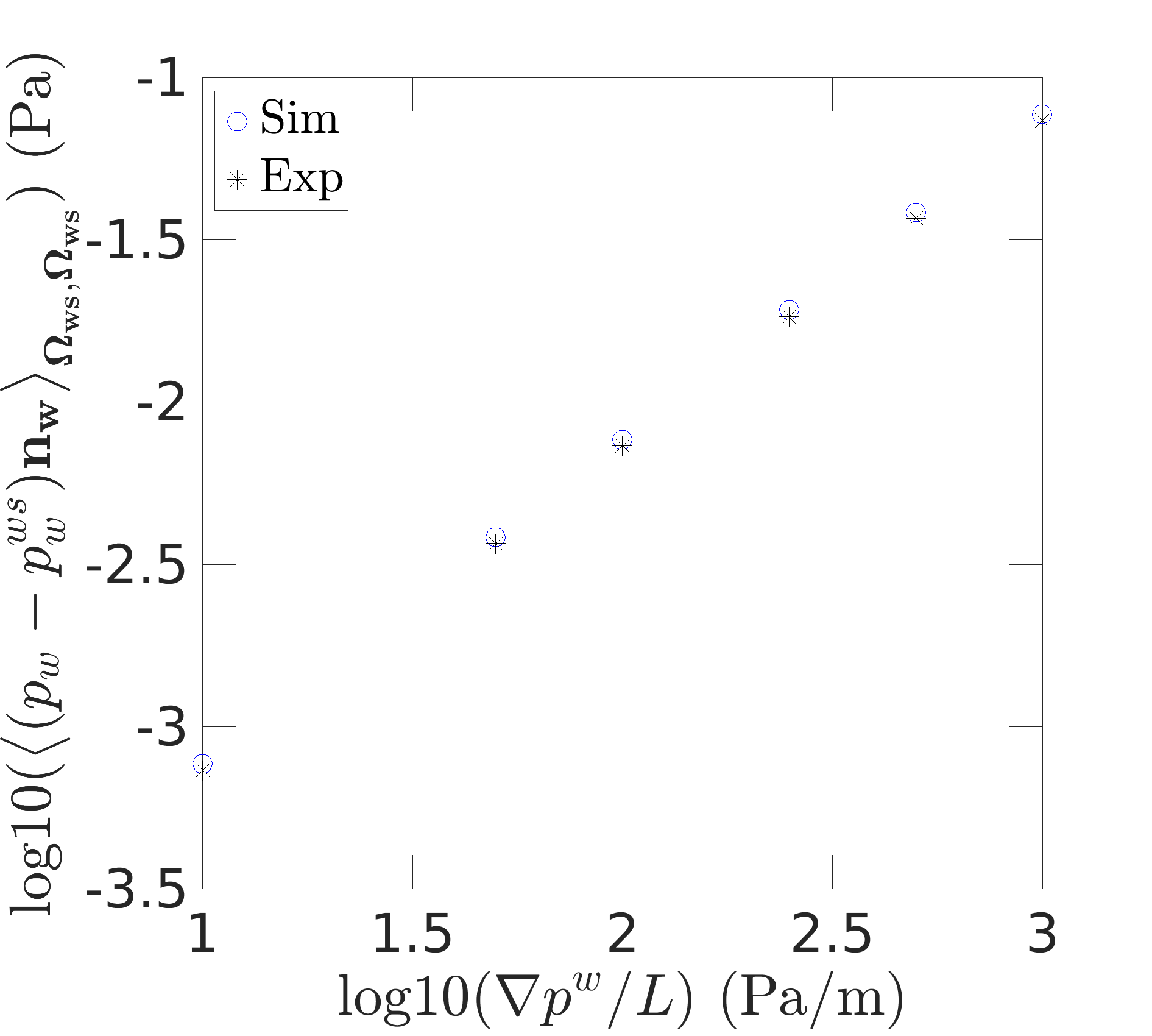}
        \caption{x-dimension}
    \end{subfigure}
    \begin{subfigure}[b]{0.329\linewidth}
        \includegraphics[width=\linewidth]{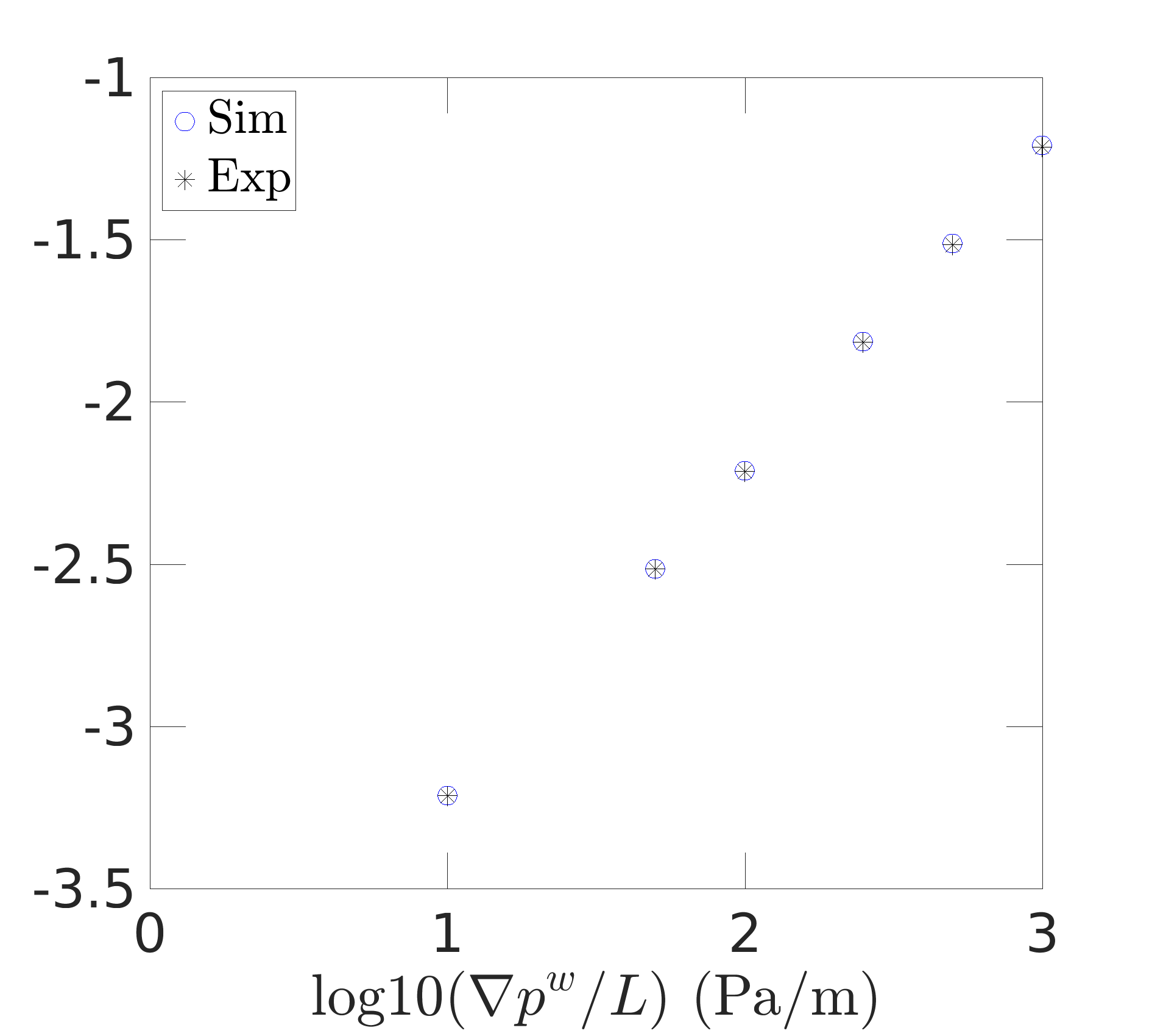}
        \caption{y-dimension}
    \end{subfigure} 
    \begin{subfigure}[b]{0.329\linewidth}
        \includegraphics[width=\linewidth]{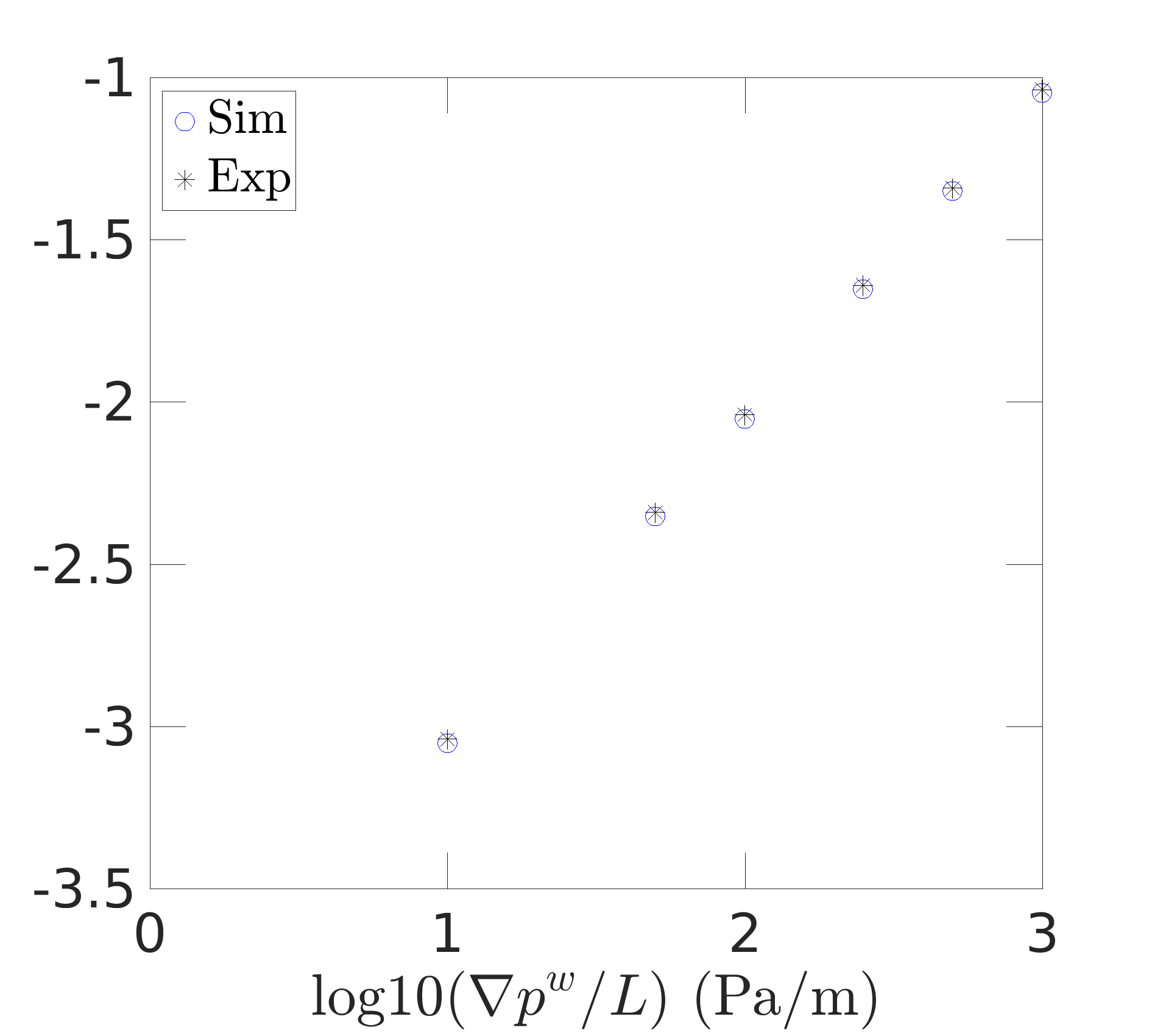}
        \caption{z-dimension}
    \end{subfigure} 
    \caption{Comparison of the simulation and expected characteristic length scales in the (a) $x$, (b) $y$, and (c) $z$ directions when the principal direction of flow was aligned with the dimension of interest.}
    \label{fig:ELLIPSE_LENGTH}
\end{figure}

Using this approximation for the characteristic length tensor to predict the pressure deviation term yielded a relative error within 5\%. The results of using this approximate length tensor versus the actual simulation data can be seen in Fig. \ref{fig:ELLIPSE_LENGTH} for the $x$, $y$, and $z$ directions of the deviation term, when the flow was in the corresponding direction. 

\begin{table*}[b!]
    \centering
    \caption{\label{tab:Zhang_Param} System and fluid parameters to reproduce results in \cite{Zhang_Prodanovic_etal_19}.}
    \begin{ruledtabular}
    \begin{tabular}{lcccc}
        Parameter \hspace{1.5cm} & \hspace{0.5cm} Fluid \#1 \hspace{0.5cm} & \hspace{0.5cm} Fluid \#2 \hspace{0.5cm} & \hspace{0.5cm} Fluid \#3 \hspace{0.5cm} & \hspace{0.5cm} Fluid \#4 \hspace{0.5cm} \\
        \hline \\
        $\Hat{\mu}_0$ (Pa$\cdot$s) & 3.52 & 3.52 & 3.52 & 3.52 \\
        $\Hat{\mu}_\infty$ (Pa$\cdot$s) & 0.001 & 0.001 & 0.1 & 0.001 \\
        $m$ (s) & 10.1 & 10.1 & 10.1 & 1 \\
        $n$ (-) & 0.62 & 0.4 & 0.62 & 0.62 \\
        $\beta$ (m$^{-1}$) &  $9.10\times 10^2$ & $8.97\times 10^2$ & $9.12\times 10^2$ & $8.81\times 10^2$ \\
        $\gka$ (-) &  1.68 & 1.70 & 1.65 & 1.70 \\
        $Re_c$ (-) &  2.07 & 2.10 & 2.07 & 2.14 \\
    \end{tabular}
    \end{ruledtabular}
\end{table*}

\subsection{Rough Slit}

For the final validation case, the results presented in \cite{Zhang_Prodanovic_etal_19} for simulation data for non-Newtonian fluid flow in a rough fracture were used. The fluid and system properties used to reconstruct the results are provided in Table \ref{tab:Zhang_Param}. To reproduce this data, the same methods used in \cite{Zhang_Prodanovic_etal_19} were used here, and then the model and {\em a priori} parameters proposed from \Eqns{RES_FORM} {MR} were used for comparison. The Newtonian intrinsic permeability for the medium was $\hat{\gkk}_N = 9.35\times 10^{-9}$ m$^2$. The change in pressure-per-length was calculated from 
\begin{equation} 
-\frac{\gkD P}{L} = \frac{\Hat{\mu}_\text{pm}}{\Hat{\gkk}_N}q + \beta\rho q^2\;,
\label{eq:RD_ZDP}
\end{equation}
where $\hat{\mu}_\text{pm}$ is called the ``porous medium" or effective viscosity and $\gkb$ is known as the intertial permeability. The porous medium viscosity is given by
\begin{equation}
\hat{\mu}_\text{pm} = \hat{\mu}_\infty + \frac{\hat{\mu}_0 - \hat{\mu}_\infty}{1 + m^n\lrp{\dot{\gkg}_\text{pm}}^n}\;,
\end{equation}
where $\dot{\gkg}_\text{pm}$ is the ``porous medium" shear rate, which is the shear rate which would be required to generate the effective viscosity. This ``porous medium" shear rate is calculated using the shift factor $\gka$ by the following equation,
\begin{equation}
\dot{\gkg}_\text{pm} = \gka \frac{q}{\sqrt{\Hat{\gkk}_N\gke^3}}\;.
\end{equation}
Inputting the parameters from Table \ref{tab:Zhang_Param} into these equations yields the data published in \cite{Zhang_Prodanovic_etal_19}. The critical Reynolds number $Re_c$ was defined in \cite{Zhang_Prodanovic_etal_19} as the Reynolds number at which the laminar term, defined as $\frac{\hat{\mu}_\text{pm}}{\hat{\gkk}_N}q$ in \Eqn {RD_ZDP}, contributes to 95\% of the measured pressure drop per length. The {\em a priori} relationship to predict the pressure drop for a non-Newtonian fluid flowing in a porous medium was developed for laminar flow, thus it should be the case that at $Re_c$ for each fluid, this relationship should predict approximately 95\% of the pressure drop in the medium; however, it is possible that the calculation of the original shift factor included some transition behavior, and that this was misinterpreted while fitting the inertial permeability term. 

\begin{figure}[t!]
    \centering
    \begin{subfigure}[b]{0.49\linewidth}
        \includegraphics[width=\linewidth]{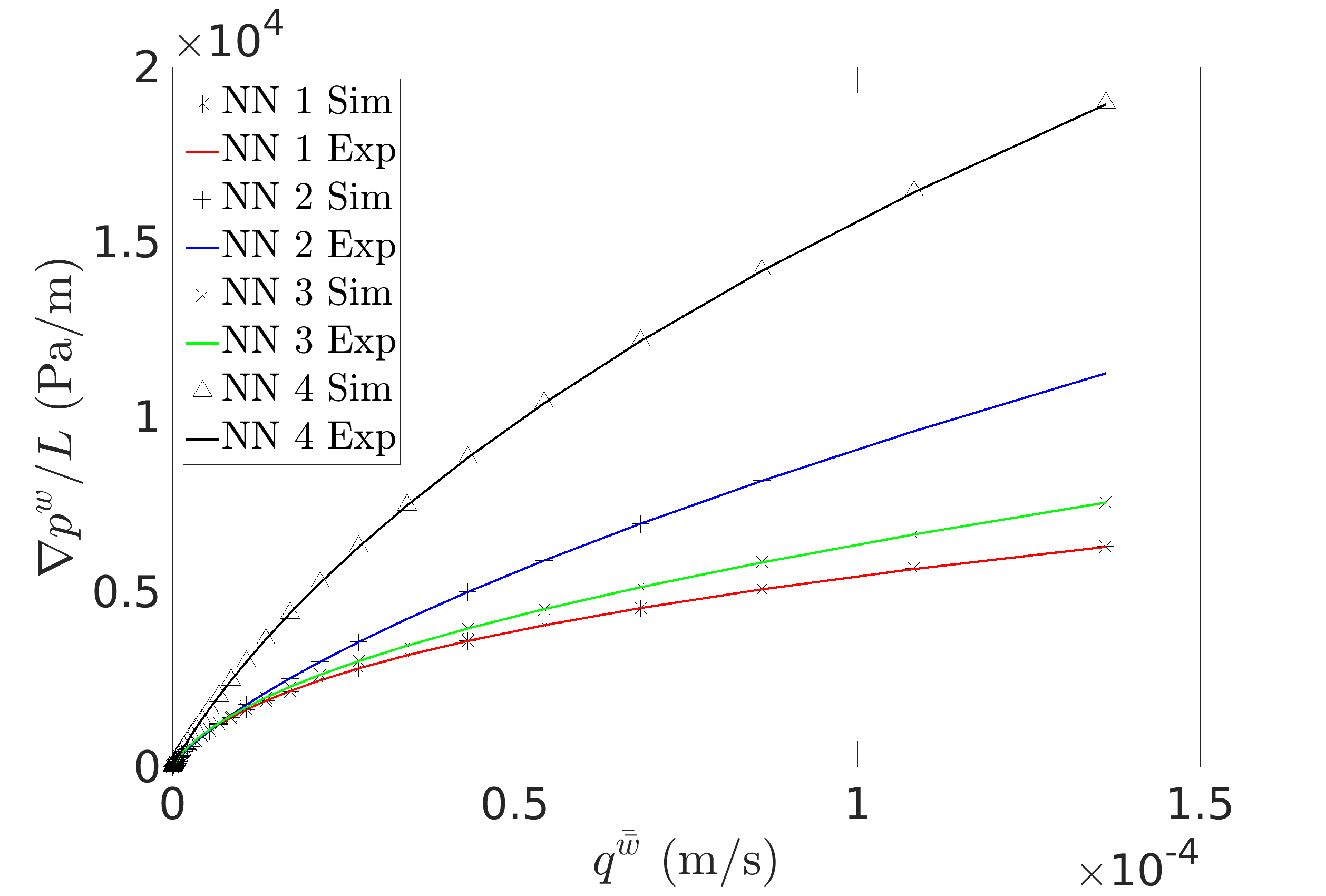}
        \caption{Laminar flow.}
    \end{subfigure}
    \begin{subfigure}[b]{0.49\linewidth}
        \includegraphics[width=\linewidth]{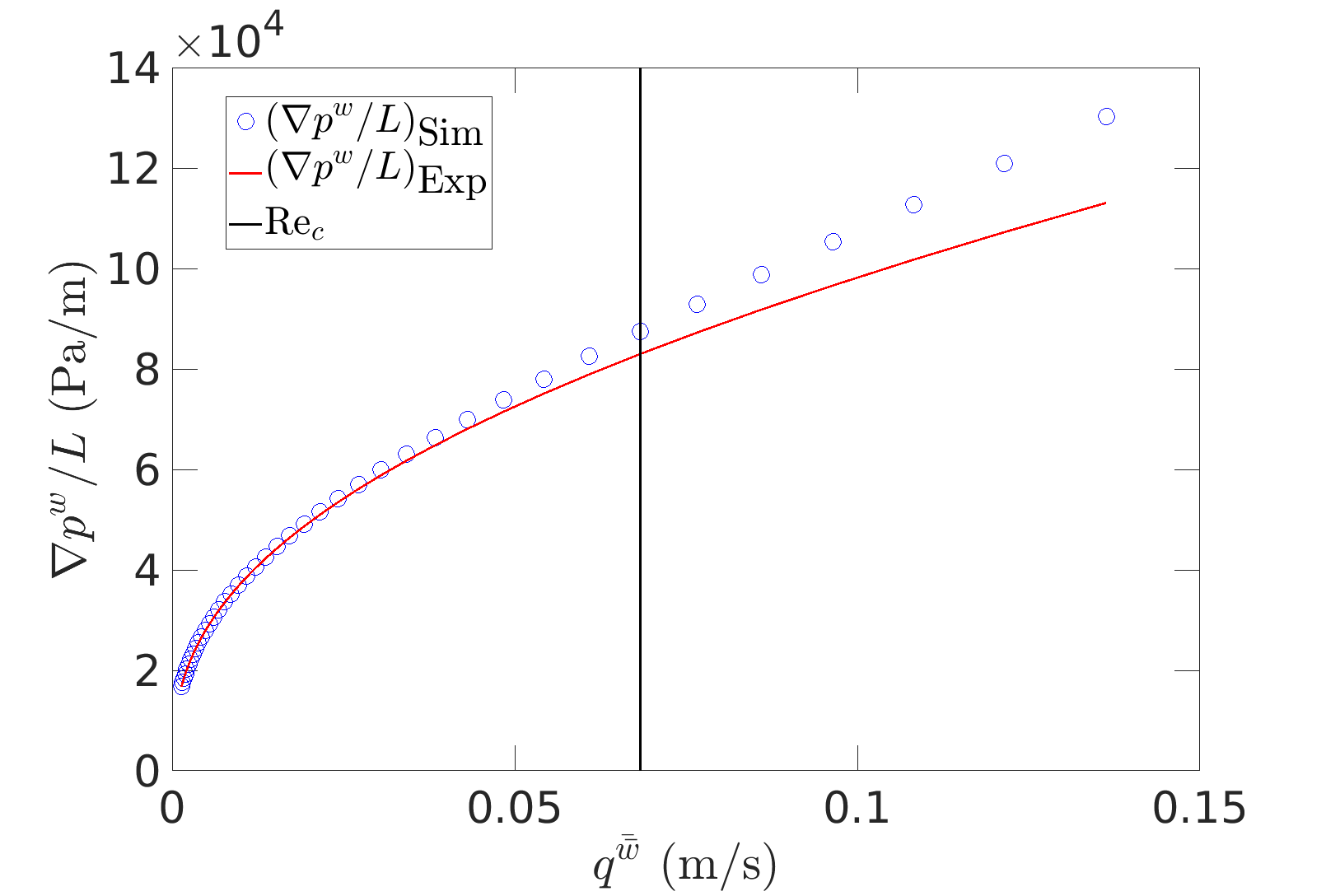}
        \caption{Onset of the transition flow regime.}
    \end{subfigure} 
    \caption{(a) Laminar comparison of simulated and expected change in pressure for all non-Newtonian fluids simulated in \cite{Zhang_Prodanovic_etal_19}. (b) Illustration of the onset of the transition flow regime for Fluid \#1, comparing actual data and expected laminar model.}
    \label{fig:ZHANG_COMP}
\end{figure}

Data from \cite{Zhang_Prodanovic_etal_19} was reproduced using \Eqn{RD_ZDP}, then $\hsR^w_\infty$ and $\hsR^w_0$ were calculated using \Eqn{RI0}, and the term $M$ was fitted for one single fluid case at a very low flow rate using \Eqn{RES_FORM}. Finally, a specific interfacial area was calculated using \Eqn{MR}. These parameters can be found in Table \ref{tab:Resistance_Param}. Using these parameters fitted using one single set of fluid data, Fig. \ref{fig:ZHANG_COMP}(a) was produced comparing the simulation data for each fluid, calculated from \Eqn{RD_ZDP}, to the expected data based on the \textit{a priori} model using \Eqn{RES_FORM}. Fig. \ref{fig:ZHANG_COMP}(b) extends the data for one of these fluids out to the critical Reynolds number, showcasing how this model may be used to determine the onset of transition flow behavior. For this case the \textit{a priori} laminar model predicts approximately 95\% of the simulation results calculated from \Eqn{RD_ZDP} as expected. 

\begin{table*}[b!]
    \centering
    \caption{\label{tab:Resistance_Param} System parameters for \Eqn{RES_FORM} produced by fitting to data from \cite{Zhang_Prodanovic_etal_19}. } 
    \begin{ruledtabular}
    \begin{tabular}{lcccc}
        Parameter \hspace{1.5cm} & \hspace{0.5cm} Fluid \#1 \hspace{0.5cm} & \hspace{0.5cm} Fluid \#2 \hspace{0.5cm} & \hspace{0.5cm} Fluid \#3 \hspace{0.5cm} & \hspace{0.5cm} Fluid \#4 \hspace{0.5cm} \\
        \hline \\
        $\hsR^w_0$ (Pa$\cdot$s/m$^2$) & 3.57$\times 10^8$ & 3.57$\times 10^8$ & 3.57$\times 10^8$ & 3.57$\times 10^8$ \\
        $\hsR^w_\infty$ (Pa$\cdot$s/m$^2$) & 1.07$\times10^5 $ & 1.07$\times10^5 $ & 1.07$\times10^8 $ & 1.07$\times10^5 $ \\
        $M$ (s/m) & 1.76$\times 10^5$ & 1.76$\times 10^5$ & 1.76$\times 10^5$ & 1.74$\times 10^4$ \\
        $N$ (-) & 0.62 & 0.4 & 0.62 & 0.62 \\
        $\ews$ (m$^{2}$/m$^3$) &  6150 & 6150 & 6150 & 6150 \\
    \end{tabular}
    \end{ruledtabular}
\end{table*}

One complication with comparing the published results to predictions based upon \Eqns{RES_FORM} {MR} is that the resistance model requires the specific solid-fluid interfacial area $\ews$, which was not provided for this set of simulations. This may be considered a further advantage of the non-Newtonian resistance model, as the results of a non-Newtonian flow experiment can be used to compute $\ews$, rather than requiring {\em in-situ} sensing techniques such as micro-CT. For the case of the slit in \cite{Zhang_Prodanovic_etal_19}, $\ews$ was fit for one of the non-Newtonian fluids, and then this was used in the rest of the fluid cases. The properties for this media may be found in \cite{Karpyn_Grader_etal_07}, where $\ews$ is reported as being approximately $12,300$ m$^2$/m$^3$, or twice the value predicted here. However, in \cite{Crandall_Bromhal_etal_10}, it was shown that many properties of this rough fracture may change dramatically based on the discretization of the mesh and the scheme used to digitize the media. To elucidate what ranges of values may be expected, media microCT images were downloaded from the Digital Rocks Portal \cite{Prodanovic_Esteva_etal_15}, these images were converted into an OBJ format using Fiji \cite{Schindelin_Carreras_etal_2012}, and the snappyHexMesh utility in OpenFOAM was used to generate a digitized mesh. The surface area of the rough fracture was calculated for a range of discretizations near the level reported in \cite{Zhang_Prodanovic_etal_19}, which had 92,458 grid blocks. It was found that the $\ews$ predicted was within the range calculated from the digitized media.  

\section{Concluding Remarks}

Non-Newtonian fluids appear regularly in many natural and industrial processes, but they are usually modeled as if they were Newtonian fluids whose viscosity changes uniformly with flow rate. It has been shown here by comparison of traditional parameters to physically significant macroscale averages that such traditional models are inadequate when attempting to predict non-Newtonian flow, and that this has lead to misunderstanding of physical macroscale data. 

A new theory of non-Newtonian laminar flow was developed here using the thermodynamically constrained averaging theory, which was based on averaging from the microscale to the macroscale. It was found by carrying out this averaging that during laminar flow the average viscosity at the fluid-solid interface is the significant macroscale parameter that impacts flow, rather than the bulk fluid viscosity or some other effective viscosity parameter. This averaging was validated for a set of parallel slits, as well as for a set of packed spheres and a set of ellipsoids. If the specific interfacial area of the system, the porosity, rheological parameters, and possible compression fluctuation forces are known, then it was shown that the average viscosity and shear rate at the fluid-solid interface can be calculated {\em a priori} for a given pressure drop. A relationship was proposed to model the fluctuation compression forces for a system to allow for \textit{a priori} prediction of the interfacial viscosity for complex isotropic and anisotropic systems. A non-Newtonian resistance model was proposed which allowed the flow rate for a non-Newtonian fluid to be predicted from fluid and system parameters for isotropic and anisotropic systems, which have generally been neglected in the non-Newtonian literature. 

This model was used to predict the flow through a set of parallel slits, for which the flow rate-pressure drop relationship is known analytically \cite{Sochi_15}, as well as to predict the flow through a rough fracture for which simulation results have been previously published \cite{Zhang_Prodanovic_etal_19}. It was also shown how this non-Newtonian laminar flow model could be used to observe the onset of transition region flow behavior, which has been difficult to determine in the past. 

Additionally, a model to describe flow of a non-Newtonian fluid in an anisotropic media was developed which required calculation of one single characteristic eigenvalue, while using other parameters that can be easily calculated using flow of a Newtonian fluid through the system, to be predictive. The results of this work can be used to model non-Newtonian flow more accurately than current approaches.

\begin{acknowledgements}
This work was supported by National Science Foundation Grant 1604314, and Army Research Office grant W911NF1920270. William G. Gray provided useful discussion related to this work.
\end{acknowledgements}

\end{document}